\documentclass[aps,prb,reprint,superscriptaddress]{revtex4-1}

\usepackage{graphicx}
\usepackage{amssymb,amsmath}
\usepackage{bm}
\usepackage{esint}

\begin{document}

\title{Elastic constants of stressed and unstressed materials in the phase field crystal model}
\author{Zi-Le Wang}
\affiliation{College of Chemistry and Molecular Engineering, Peking University,
  Beijing 100871, China}
\author{Zhi-Feng Huang}
\email{huang@wayne.edu}
\affiliation{Department of Physics and Astronomy, Wayne State University,
  Detroit, Michigan 48201, USA}
\author{Zhirong Liu}
\email{LiuZhiRong@pku.edu.cn}
\affiliation{College of Chemistry and Molecular Engineering, Peking University,
Beijing 100871, China}
\affiliation{Center for Quantitative Biology, and Beijing National Laboratory for
  Molecular Sciences (BNLMS), Peking University, Beijing 100871, China}

\date{\today}

\begin{abstract}
A general procedure is developed to investigate the elastic response and calculate the
elastic constants of stressed and unstressed materials through continuum field modeling,
particularly the phase field crystal (PFC) models. It is found that for a complete description
of system response to elastic deformation, the variations of all the quantities of lattice
wave vectors, their density amplitudes (including the corresponding anisotropic variation
and degeneracy breaking), the average atomic density, and system volume should be incorporated.
The quantitative and qualitative results of elastic constant calculations highly depend on the
physical interpretation of the density field used in the model, and also importantly, on the
intrinsic pressure that usually pre-exists in the model system. A formulation based on
thermodynamics is constructed to account for the effects caused by constant pre-existing stress
during the homogeneous elastic deformation, through the introducing of a generalized Gibbs
free energy and an effective finite strain tensor used for determining the elastic constants.
The elastic properties of both solid and liquid states can be well produced by this unified
approach, as demonstrated by an analysis for the liquid state and numerical evaluations for
the bcc solid phase. The numerical calculations of bcc elastic constants and Poisson's
ratio through this method generate results that are consistent with experimental conditions,
and better match the data of bcc Fe given by molecular dynamics simulations as compared to
previous work. The general theory developed here is applicable to the study of different types
of stressed or unstressed material systems under elastic deformation.
\end{abstract}

\maketitle

\section{Introduction}

The phase field crystal (PFC) approach is an effective methodology emerging in recent years
which describes the formation and dynamics of complex spatial structures and patterns with
atomic resolution. \cite{prl245701,prl225504,prb64107,pre21605,ap665} 
Compared to conventional phase field models for microstructure evolution,\cite{prb6119,arms113} 
the PFC method provides an explicit description of the atomic density distribution, 
with which the coupling between microscopic and mesoscopic scales is more straightforwardly
incorporated. It also naturally incorporates system elastic energy and effects of topological
defects (a feature that is otherwise much more difficult to implement in conventional phase
field models \cite{prb35429}), and has been widely used to study a variety of phenomena in
condensed matter physics and materials science, such as grain boundary energies, structures,
and dynamics,\cite{prl245701,pre51605,prb184110,prl255501}
crystal-liquid interfaces,\cite{prb184107,pre31602}
crystal growth,\cite{prl35702,pre012405}
plasticity and dislocation dynamics,\cite{pre51605,prl15502}
ferromagnetics and ferroelectrics,\cite{prb184109}
order-disorder transition,\cite{pre22105} among many others. Several PFC-type models
have been developed to successfully model a number of crystal structures and ordered patterns.
\cite{prl045702,jpc205402,jpc364102,prm060801,pre53305,npj15013,prl35501,prl205502}
For example, in two-dimensional systems a PFC model featuring three competing length scales
has been constructed to produce all five Bravais lattices and some complex structures such as
honeycomb, kagome, dimer, and some hybrid ordered phases.\cite{prl35501}

Although elasticity is one of the focus points of the PFC modeling from the very beginning,
\cite{prl245701,pre51605} the issue of how to accurately calculate elastic constants
in the PFC models is still not well resolved. In most studies, elastic constants
were calculated from the variations of free energy density caused by various types of
strains at a constant average atomic density ($\bar{\rho}$). \cite{prl245701,pre51605,
  prl35501,prl205502,prb125408,pre61601,pre11602,prb214105} 
However, as pointed out by Pisutha-Arnond {\it et al.},\cite{prb14103} this procedure is
inconsistent with the fact that $\bar{\rho}$ is actually affected by the strain imposed;
e.g., $\bar{\rho}$ gets smaller under a tensile strain while it gets larger under a
compressive strain. Secondly, the amplitudes of atom density distribution were usually
assumed to be unvaried under strain, which greatly facilitates the analyses. 
But this is not necessarily valid either, since an anisotropic deformation not only
causes the variation of amplitudes, but also breaks the degeneracy of amplitudes in
the one-mode approximation.\cite{prb214105} Lastly, the equilibrium phases in the PFC
models are usually optimized under a constant $\bar{\rho}$, which may be highly stressed
(i.e., under high pressure). For example, for PFC model the estimated pressure at the
liquid-solid coexistence is as high as $1.8\times 10^6$ atm under a set of parameters
for iron (Fe).\cite{prb14103} For stressed materials, there are various sets of elastic
coefficients or constants that differ from each other in the thermoelasticity theory.
\cite{pr776,prb423,jasa348} The applicability of these different coefficients in the
PFC models is confusing and needs further clarification.

In this paper, we focus on how elastic constants should be calculated accurately within
the PFC framework, and the formulation developed is applicable to other types of
coarse-graining continuum field models. Our study shows that to produce reasonable
results consistent with conditions of real materials, the deformation-induced changes
of average atomic density, volume, and amplitudes of density waves, in addition to
the density wave vectors themselves, should be all considered. Particular attention
needs to be paid to the representation of order-parameter density field in the model.
Although various forms of atomic density field are equally valid in the PFC modeling,
different schemes should be used to describe their variations under strain, which
would affect the outcomes of elastic constant calculations including both the quantitative
values and their change with varying average density. Another important factor is
the pre-existing nonzero pressure in the undeformed state of PFC (particularly
with the absence of the linear term in the PFC free energy functional). This requires
the constructing of a new thermodynamic formulation to incorporate the pre-stressed
state in both solid and liquid phases, so that a proper thermodynamic definition or
calculation of elastic constants can be obtained, through either a generalized Gibbs
free energy (as in $NPT$ ensemble) or a new finite strain tensor if using the Helmholtz
free energy. This set of elastic constants defined through thermodynamic potential
has complete Voigt symmetry, and is equivalent to the symmetric combination of $B$
elastic coefficients \cite{pr776} determined through the stress-strain relation.
The validity of our approach is demonstrated in both analytic and numerical examples,
including a liquid-phase elastic analysis and numerical calculations of elastic
constants of bcc Fe that are compared to results of molecular dynamics (MD) simulations
and previous PFC studies.

The rest of this paper is organized as follows. 
In Sec.~\ref{sec:theory}, a general theoretical framework is developed to describe
the elastic response of any specific phase and to calculate elastic constants
under stressed condition. In Sec.~\ref{sec:liquid}, liquid is adopted as a simple
analytic example to test the validity of various definitions of elastic constants.
It is shown that some previous definitions would be improper, and only the result
generated from this approach is consistent with the property of liquid.
In Sec.~\ref{sec:results}, the theory developed is applied to numerically analyze
the elastic properties of bcc Fe, showing different consequences of various algorithms
and options and the effectiveness of our method. Finally, we discuss and summarize
our results in Sec.~\ref{sec:discussion} and Sec.~\ref{sec:summary}.

\section{Theory and Model}
\label{sec:theory}

\subsection{PFC model}

We consider the simplest one-mode PFC model \cite{prl245701,pre51605,prb64107} with the
free energy functional given by
\begin{equation}
  \mathcal{F}[\phi(\mathbf{r})]=\int_V d\mathbf{r}
  \left\{a\phi + \frac{\phi}{2} \left[ b+\lambda\left( \nabla^2 + q_0^2 \right)^2 \right] \phi
  + \frac{g}{4} \phi^4 \right\},
\label{PFC-1}
\end{equation}
where $\phi(\mathbf{r})$ is the atom number density difference with respect to a
uniform reference-state density $\rho_0$, i.e., 
\begin{equation}
\phi(\mathbf{r})=\Delta\rho(\mathbf{r})=\rho(\mathbf{r})-\rho_0,
\label{PFC-2}
\end{equation}
and $a$, $b$, $\lambda$, $q_0$, and $g$ are phenomenological parameters. 
In principle, the cubic term $\phi^3$ should be included in the free energy expansion,
\cite{prb64107} but it can always be removed by properly choosing the reference-state
$\rho_0$.\cite{pre31602} The linear term $a\phi$ was usually ignored in previous modeling
since the corresponding integration in $\mathcal{F}$ gives a trivial term of $aV\bar{\phi}$ 
that has no influence on the phase stability given constant volume $V$ and average density
difference $\bar{\phi}$. However, it plays an essential role on determining the pressure
of the system and the calculation of elastic constants, as will be shown below; thus we
explicitly include it in the above free energy functional.

It is convenient to rescale the free energy functional to a dimensionless form by
setting \cite{pre51605} $q_0\mathbf{r}\rightarrow \mathbf{r}$,
$\sqrt{g/\lambda q_0^4}\phi \rightarrow \psi$, and $(g/\lambda^2 q_0^5)\mathcal{F}
\rightarrow \mathcal{F}$ in three-dimensional (3D) systems, leading to
\begin{equation}
  \mathcal{F}[\psi(\mathbf{r})]=\int_V d\mathbf{r}
  \left\{\alpha\psi + \frac{\psi}{2} \left[ -\epsilon+\left( \nabla^2+1\right)^2 \right] \psi
  +\frac{1}{4}\psi^4 \right\},
\label{PFC-4}
\end{equation}
with the dimensionless parameters $\alpha=(a/q_0^6)\sqrt{g/\lambda^3}$ and
$\epsilon=-b/\lambda q_0^4$. Here parameter $\epsilon$ is generally considered to change
with temperature $T$ and measure the distance from liquid-solid transition.
\cite{pre51605,prb64107}

In some PFC studies, \cite{pre31602,pre21605,prb64107} the dimensionless density variation
field $n$, defined as
\begin{equation}
n(\mathbf{r})=\frac{\Delta\rho(\mathbf{r})}{\rho_0}=\frac{\rho(\mathbf{r})-\rho_0}{\rho_0}, 
\label{PFC-6}
\end{equation}
is used instead of $\psi$ or $\phi$ in the free energy functional. These density fields
$\rho$, $\phi$, $\psi$, and $n$ are equally valid in describing the atomic density
distribution in the PFC model, but their variations caused by strain are different. 
When a strain is applied to the system, the total number of particles,
\begin{equation}
N=\int_V \rho(\mathbf{r}) d\mathbf{r} = \bar{\rho}V,
\label{PFC-7}
\end{equation}
remains constant during the elastic deformation. Thus the relation between the average
density of the strained system and that of the unstrained one is given by
\begin{equation}
\bar{\rho}_{\rm strained}=\frac{V_{\rm unstrained}}{V_{\rm strained}} \bar{\rho}_{\rm unstrained},
\label{PFC-8}
\end{equation}
where $V_{\rm unstrained}$ and $V_{\rm strained}$ are the undeformed and deformed volumes,
respectively. From Eq.~(\ref{PFC-8}) and the definitions of $n$, $\phi$, and $\psi$,
we have 
\begin{eqnarray}
&&  \bar{n}_{\rm strained}=\frac{V_{\rm unstrained}}{V_{\rm strained}}
  \left(\bar{n}_{\rm unstrained}+1\right)-1,
\label{PFC-9}\\
&&  \bar{\phi}_{\rm strained}=\frac{V_{\rm unstrained}}{V_{\rm strained}}
  \left(\bar{\phi}_{\rm unstrained}+\rho_0\right)-\rho_0,
\label{PFC-10}\\
&&  \bar{\psi}_{\rm strained}=\frac{V_{\rm unstrained}}{V_{\rm strained}}
  \left(\bar{\psi}_{\rm unstrained}+\tilde{\rho}_0\right)-\tilde{\rho}_0,
\label{PFC-11}
\end{eqnarray}
where
\begin{equation}
\tilde{\rho}_0 = \sqrt{\frac{g}{\lambda q_0^4}} \rho_0.
\label{PFC-12}
\end{equation}
It is worth pointing out that $\rho_0$ (or $\tilde{\rho}_0$) is not needed in
examining the stability and dynamics of various phases,
but it plays a non-negligible role on the elastic constant
calculations as will be demonstrated in Sec.~\ref{sec:results}. 

In a crystalline state, $\psi$ can be expressed in terms of Fourier components, i.e.,
\begin{equation}
\psi(\mathbf{r})=\bar{\psi}+\sum_{\mathbf{K}} A_{\mathbf{K}}e^{i\mathbf{K}\cdot\mathbf{r}},
\label{PFC-13}
\end{equation}
where $\mathbf{K}$ is the nonzero reciprocal lattice vector and $A_{\mathbf{K}}$ is
the corresponding Fourier-component amplitude with $A_{-\mathbf{K}}=A_{\mathbf{K}}^*$. 
Substituting Eq.~(\ref{PFC-13}) into Eq.~(\ref{PFC-4}), the free energy functional can
be written as the form $\mathcal{F}(\{A_{\mathbf{K}}\}, \{\mathbf{K}\}, \bar{\psi}, V)$.
The equilibrium undeformed state is determined by minimizing the free energy:
\begin{equation}
F_{\rm unstrained}=\min_{\left\{A_{\mathbf{K}}\right\}, \left\{\mathbf{K}\right\}} 
\mathcal{F}\left( \left\{A_{\mathbf{K}}\right\}, \left\{\mathbf{K}\right\}, \bar{\psi}, V \right),
\label{PFC-15}
\end{equation}
under the condition of fixed $\bar{\psi}$ and $V$ (and the resulting total particle
number $N$). Note that a state with any value of $\bar{\psi}$ can be chosen as the
initial undeformed state which, however, is not necessarily unstressed due to, e.g.,
a pre-existing pressure $P_0$ in the system (see below). When a specific lattice symmetry,
e.g., the body-centered cubic (bcc) phase, is considered, there is only one free parameter
in specifying $\{\mathbf{K}\}$, usually chosen as the smallest length of $\mathbf{K}$,
i.e., $|\mathbf{K}_{110}| =|\mathbf{K}_{1\bar{1}0}|=|\mathbf{K}_{101}|=|\mathbf{K}_{10\bar{1}}|
=|\mathbf{K}_{011}|=|\mathbf{K}_{01\bar{1}}| \equiv q_0$ for the first mode of bcc.
In addition, usually $A_{\mathbf{K}}$ with low-index $\mathbf{K}$ has much larger magnitude;
hence a few-mode approximation can be adopted to simplify analysis. For example, in a
one-mode approximation of bcc phase, only the first group of $\mathbf{K}$ are considered 
in the analysis, and their amplitudes $A_{\mathbf{K}}$ are identical due to lattice symmetry:
\begin{equation}
A_{110} =A_{1\bar{1}0}=A_{101}=A_{10\bar{1}}=A_{011}=A_{01\bar{1}} \equiv A. 
\label{PFC-17}
\end{equation}
The corresponding free energy functional then becomes $\mathcal{F}( A, q_0, \bar{\psi}, V)$.
However, when a uniaxial or shear strain is applied, the lattice would be
distorted anisotropically. Thus Eq.~(\ref{PFC-17}) is no longer satisfied and a single
amplitude $A$ is not sufficient in the description.

\subsection{Strain tensors and elastic response}
\label{sec:strain}

A homogeneous elastic strain upon a crystalline state can be measured by a tensor of
displacement gradients, ${\bm \nabla} \mathbf{u}=\left\{ u_{ij}\right\}$, 
which transforms any lattice vector in an initial undeformed state ($\mathbf{R}$) 
to that in a deformed state ($\mathbf{R}^{\prime}$):
\begin{equation}
d\mathbf{R}^{\prime}=\left(\mathbf{I}+{\bm \nabla} \mathbf{u}\right) \cdot d\mathbf{R},
\label{Strain-1}
\end{equation}
where $\mathbf{I}$ is the unit vector. The displacement gradient tensor ${\bm \nabla} \mathbf{u}$
can be separated into two parts, i.e., $u_{ij} = \varepsilon_{ij} + \omega_{ij}$.
The infinitesimal strain tensor $\boldsymbol{\varepsilon}$ (i.e., Cauchy's strain tensor 
or linear strain tensor) is defined as the symmetric components of ${\bm \nabla} \mathbf{u}$:
\begin{equation}
\varepsilon_{ij}=\frac{1}{2} \left( u_{ij}+u_{ji}\right),
\label{Strain-2}
\end{equation}
while the antisymmetric part (i.e., the rotational tensor), 
\begin{equation}
\omega_{ij}=\frac{1}{2} \left( u_{ij}-u_{ji}\right),
\label{Strain-3}
\end{equation}
measures pure rotation and does not affect the system energy due to rotational invariance
of the system. Thus in this study we neglect $\omega_{ij}$ to facilitate the analysis. 
For finite strain, the definition of finite strain tensors is essential to the nonlinear
elasticity theory.\cite{prb214105} A widely adopted finite strain tensor is the Lagrangian
strain tensor $\mathbf{E}$ (i.e., the Green-Lagrangian strain tensor), defined as 
\begin{equation}
E_{ij}=\frac{1}{2} \left( u_{ij}+u_{ji}+u_{ki}u_{kj} \right),
\label{Strain-4}
\end{equation}
where the Einstein summation convention for repeated indices is used.
Note that all the strain tensors defined in this work are measured with respect to
the initial state ($\mathbf{R}$) which could be either unstressed or stressed.
For unstressed systems, the difference between infinitesimal and finite strain tensors 
is unimportant in the linear elasticity theory. For stressed systems, however, the
difference is significant even for linear elasticity,\cite{pr776,prb14103} 
which should be treated cautiously as will be demonstrated below.

A strain changes the lattice vectors $\{ \mathbf{R} \}$ and distorts the unit cell of
a crystalline phase. The reciprocal lattice vectors $\{ \mathbf{K} \}$ are also changed
accordingly. From $\mathbf{K}^{\rm (strained)} \cdot \mathbf{R}^{\rm (strained)} =
\mathbf{K}^{\rm (unstrained)} \cdot \mathbf{R}^{\rm (unstrained)}$, the strained reciprocal
lattice vectors are given by
\begin{equation}
  \mathbf{K}^{\rm (strained)} = \left(\mathbf{I}+\boldsymbol{\varepsilon}\right)^{-1} \cdot
  \mathbf{K}^{\rm (unstrained)}
\label{Strain-5}
\end{equation}
under elastic deformation, where $\mathbf{K}^{\rm (unstrained)}$ are the equilibrium
$\mathbf{K}$ obtained from the free energy minimization [see Eq.~(\ref{PFC-15})]. 
Here we have replaced ${\bm \nabla} \mathbf{u}$ by $\boldsymbol{\varepsilon}$
in lattice transformation, given the lack of energy contribution from $\omega_{ij}$
in rotationally invariant systems. Under a strain $\boldsymbol{\varepsilon}$, the
volume of the system changes as
\begin{equation}
V_{\rm strained} = \det \left[\mathbf{I}+\boldsymbol{\varepsilon}\right] V_{\rm unstrained}.
\label{Strain-6}
\end{equation}
The change of volume leads to the variation of the average particle density as
shown in Eqs.~(\ref{PFC-8})--(\ref{PFC-11}) due to the conservation of particle number. 
With the constraint of $\{ \mathbf{K} \}$, $V$, and $\bar{\psi}$, the only left variables
for the free energy functional determining the system relaxation in elastic response
are $\{ A_{\mathbf{K}} \}$. It is important to note that atomic relaxation within
a unit cell after the deformation is accompanied by the variation of $\{ A_{\mathbf{K}} \}$.
Therefore, the free energy of the strained (deformed) state is written as
\begin{eqnarray}
&& F_{\rm strained}= \nonumber\\
&& \min_{\left\{A_{\mathbf{K}}\right\}} 
\mathcal{F}\left( \left\{A_{\mathbf{K}}\right\}, \left\{\mathbf{K}^{\rm (strained)}\right\},
\bar{\psi}_{\rm strained}, V_{\rm strained} \right),
\label{Strain-7}
\end{eqnarray}
where $\mathbf{K}^{\rm (strained)}$, $\bar{\psi}_{\rm strained}$, and $V_{\rm strained}$ are
determined by Eqs.~(\ref{Strain-5}), (\ref{PFC-11}), and (\ref{Strain-6}), respectively. 
In previous studies, various incomplete schemes were used in describing the free
energy response under strain. The overwhelming majority of studies considered only
the variation of $\mathbf{K}$ while ignoring the change of average density $\bar{\psi}$. 
\cite{prl245701,pre51605,prl35501,prl205502,prb125408,pre61601,pre11602,prb214105} 
Among them the anisotropic variation of $A_{\mathbf{K}}$ under strain was addressed only in
Ref. \onlinecite{prb214105}. Pisutha-Arnond {\it et al.} have considered the variations of
$\mathbf{K}$, $\bar{\psi}$, and $V$,\cite{prb14103} but neglected the varying of $A_{\mathbf{K}}$. 
In this study, we suggest that the variations of $\mathbf{K}$, $A_{\mathbf{K}}$, $\bar{\psi}$, 
and $V$ are all needed to properly describe the strain response in PFC models.

For liquids, the variations of $\bar{\psi}$ and $V$ are sufficient in describing the elastic
response since $A_{\mathbf{K}}=0$ for nonzero $\mathbf{K}$. Thus the elastic properties of
liquids can also be determined by the above procedure as a special case, which will be further
analyzed in Sec.~\ref{sec:liquid}.

\subsection{Elastic constants under pre-existing stress}

In the following we discuss how to calculate elastic constants from the free energy of
strained system. For solid and liquid states described above, the free energy of both
unstrained and strained systems can be written as a function of free variables in the
form $F(\boldsymbol{\varepsilon}, \bar{\psi}, V_0)$, where $\bar{\psi}$ and $V_0$ are
the average density and volume of the initial undeformed state, respectively, and
$\boldsymbol{\varepsilon}$ is the linear strain tensor characterizing the applied
strain with respect to the initial state. 
Alternately, the Lagrangian finite strain tensor $\mathbf{E}$ can be used to characterize
the strain, and the free energy can be written similarly as $F(\mathbf{E}, \bar{\psi}, V_0)$.

Contrary to popular belief, we will show below that the isothermal elastic constants
$C_{ijkl}$ are not necessarily equal to the second-order derivatives of $F$ with respect
to $\boldsymbol{\varepsilon}$, i.e., 
\begin{equation}
  C_{ijkl} \neq \frac{1}{V_0} \left. \frac{\partial^2 F}
  {\partial \varepsilon_{ij} \partial \varepsilon_{kl}} \right |_{\boldsymbol{\varepsilon}=0},
\label{Eq1-1}
\end{equation}
and neither are they necessarily equal to those with respect to $\mathbf{E}$:
\begin{equation}
  C_{ijkl} \neq \frac{1}{V_0} \left.
  \frac{\partial^2 F} {\partial E_{ij} \partial E_{kl}} \right |_{\mathbf{E}=0}.
\label{Eq1-1a}
\end{equation}
The reason lies in the fact that the initial undeformed state could be pre-stressed
(for which the strain, either $\varepsilon_{ij}$ or $E_{ij}$, is measured from the
initial stressed state). 
In the usual procedure of PFC study, the free energy of an equilibrium undeformed
state is minimized under fixed $\bar{\psi}$ and $V_0$ and the resulting fixed $N$
[see Eq.~(\ref{PFC-15})], i.e., within the $NVT$ ensemble 
[where $T$ is related to the parameter $\epsilon$ in Eq.~(\ref{PFC-4})]. 
Therefore, the free energy $F$, or $\mathcal{F}$ in the PFC models, is the Helmholtz
free energy but not the Gibbs free energy. The equilibrium pressure of the initial
undeformed state can be determined from 
\begin{eqnarray}
  P_0\left(\bar{\psi}\right)
  &=& -\frac{1}{V_0} \left. \frac{\partial F
    \left(\boldsymbol{\varepsilon}, \bar{\psi}, V_0 \right)}{\partial \varepsilon_{ii}}
  \right |_{\boldsymbol{\varepsilon}=0} \nonumber \\
  &=&-\frac{1}{V_0} \left. \frac{\partial F
    \left(\mathbf{E}, \bar{\psi}, V_0 \right)}{\partial E_{ii}} \right |_{\mathbf{E}=0}.
\label{Eq1-1b}
\end{eqnarray}
This pre-existing pressure $P_0$ is independent of $V_0$ given that $F$ is proportional
to $V_0$. When discussing within the $NPT$ ensemble, an external pressure equal to $P_0$ is
required to stabilize the whole system. Thus elastic constants $C_{ijkl}$ are equal to
$(1/V_0) {\partial^2 F}/{\partial \varepsilon_{ij} \partial \varepsilon_{kl}}$ 
or $(1/V_0) {\partial^2 F}/{\partial E_{ij} \partial E_{kl}}$ only when $P_0=0$. 
When $P_0\neq 0$ (which is usually the case in the PFC models especially when
the linear term in the free energy functional is ignored in previous studies), 
a modified formula should be used as explained in the following.

The elastic coefficients characterize how easy or difficult the system can be deformed, 
and are determined from the work required to strain the system. It is noted that the ``work''
here refers to the actual work that is performed in addition to the pre-existing expansion
or compression work done by the constant pressure $P_0$ (or by any pre-existing constant
external stress); i.e., it is the actual work under constant temperature and pressure
($NPT$ ensemble). Therefore, to calculate elastic constants we should consider the Gibbs
free energy $G$ instead of $F$ (see also Appendix \ref{sec:appen_elastic}), with
\begin{equation}
G\left(\boldsymbol{\varepsilon}, \bar{\psi}, V_0 \right)
= F\left(\boldsymbol{\varepsilon}, \bar{\psi}, V_0 \right)
+P_0\left(\bar{\psi} \right) V\left(\boldsymbol{\varepsilon}, V_0 \right),
\label{Eq2-1}
\end{equation}
where $P_0\left(\bar{\psi} \right)$ is given in Eq.~(\ref{Eq1-1b}) 
(so that we have $\partial G/\partial \varepsilon_{ii} |_{\boldsymbol{\varepsilon}=0} = 0$
as required by system stability), and $V$ is the deformed (strained) volume given in
Eq.~(\ref{Strain-6}) which can be expanded to second order of $\boldsymbol{\varepsilon}$ as
\begin{eqnarray}
&& V\left(\boldsymbol{\varepsilon}, V_0 \right) = V_0 \left | 
\begin{array}{ccc}
1+\varepsilon_{11} & \varepsilon_{12} & \varepsilon_{13} \\
\varepsilon_{21} & 1+\varepsilon_{22} & \varepsilon_{23} \\
\varepsilon_{31} & \varepsilon_{32} & 1+\varepsilon_{33}
\end{array}
\right | \nonumber \\
&&= V_0 \left ( 1 + \varepsilon_{11}+\varepsilon_{22}+\varepsilon_{33} 
+ \varepsilon_{11}\varepsilon_{22}+\varepsilon_{11}\varepsilon_{33}+\varepsilon_{22}\varepsilon_{33}
\right. \nonumber\\
&& \qquad \left. -\varepsilon_{12}\varepsilon_{21}-\varepsilon_{13}\varepsilon_{31}
-\varepsilon_{23}\varepsilon_{32} \right )
+ \mathcal{O}(\boldsymbol{\varepsilon}^3) \nonumber \\
&&= V_0 \left[ 1+ \varepsilon_{ii} + \frac{1}{2} \left( \varepsilon_{ii}\varepsilon_{jj}
  -\varepsilon_{ij}\varepsilon_{ji} \right) \right]
+ \mathcal{O}(\boldsymbol{\varepsilon}^3).
\label{Eq4-4}
\end{eqnarray}
Elastic constants are determined by [see Eq.~(\ref{Eq1-19})]
\begin{equation}
C_{ijkl} = \frac{1}{V_0} \left.
\frac{\partial^2 G\left(\boldsymbol{\varepsilon}, \bar{\psi}, V_0 \right)} 
{\partial \varepsilon_{ij} \partial \varepsilon_{kl}} \right |_{\boldsymbol{\varepsilon}=0}.
\label{Eq1-1c}
\end{equation}

Equation (\ref{Eq1-1c}) also applies to a more general case of stressed materials under
any pre-existing constant stress $\boldsymbol{\sigma}^{\rm (0)}$ (either isotropic or
anisotropic), for which $G$ is then a generalized Gibbs free energy [see Eqs.~(\ref{Eq1-17c})
and (\ref{Eq_G})]
\begin{eqnarray}
  G &=& F + \left [ P_0 - \left( \varepsilon_{ij} - \frac{1}{2}\varepsilon_{ik}\varepsilon_{kj}
  + \frac{1}{2}\varepsilon_{ij}\varepsilon_{kk} \right) \sigma^{\rm (0)}_{ij} \right ] V_0 \nonumber\\
  &=& F + \left ( P_0 - \xi_{ij} \sigma^{\rm (0)}_{ij} \right ) V_0.
\label{Eq1-15a}
\end{eqnarray}
The detailed derivation for systems under homogeneous elastic deformation is presented in
Appendix \ref{sec:appen_elastic}. Here an effective finite strain tensor ${\bm \xi}=\{\xi_{ij}\}$
has been introduced, with
\begin{equation}
  \xi_{ij}= \varepsilon_{ij}- \frac{1}{2}\varepsilon_{ik}\varepsilon_{kj}
  + \frac{1}{2}\varepsilon_{ij}\varepsilon_{kk},
\label{Eq1-20a}
\end{equation}
and $\boldsymbol{\sigma}^{\rm (0)}=\{ \sigma^{\rm (0)}_{ij} \}$ is the external stress tensor
required to equilibrate and stabilize the initial undeformed state, i.e.,
\begin{equation}
  \sigma^{\rm (0)}_{ij}=\frac{1}{V_0} \left. \frac{\partial F
    \left(\boldsymbol{\varepsilon}, \bar{\psi}, V_0 \right)}{\partial \varepsilon_{ij}}
  \right |_{\boldsymbol{\varepsilon}=0}.
\label{Eq1-17a}
\end{equation}
Note that when this stress tensor is isotropic, i.e., ${\sigma}^{\rm (0)}_{ij}=
-P_0\delta_{ij}$, the standard formula of Gibbs free energy Eq.~(\ref{Eq2-1}) can be
recovered from the generalized formulation of Eq.~(\ref{Eq1-15a}).

Given the definition of $\xi_{ij}$, it can be proved that (see Appendix \ref{sec:appen_elastic})
Eq.~(\ref{Eq1-1c}) for determining isothermal elastic constants is equivalent to
\begin{equation}
  C_{ijkl}= \frac{1}{V_0} \left. \frac{\partial^2 F}{\partial \xi_{ij} \partial \xi_{kl}}
  \right |_{{\bm \xi}=0},
\label{Eq1-26a}
\end{equation}
which is more convenient for both analytic and numerical calculations since the evaluation
of $\partial F/\partial \varepsilon_{ij}$ and $\sigma^{\rm (0)}_{ij}$ in Eq.~(\ref{Eq1-17a})
is not needed here.

\section{Analysis of liquid state}
\label{sec:liquid}

In previous work,\cite{prb14103,pr776} both infinitesimal and finite strain tensors 
($\boldsymbol{\varepsilon}$ and $\mathbf{E}$) were used in defining elastic constants
$C_{ijkl}$. In this section, we use liquid as a simple analytic example to demonstrate
that for a stressed system $C_{ijkl}$ cannot be defined as the second-order derivatives
of free energy $F$ with respective to $\boldsymbol{\varepsilon}$ or $\mathbf{E}$, but
should be defined as that to the new strain tensor $\boldsymbol{\xi}$ as given in
Eq.~(\ref{Eq1-26a}). This study of liquid state is motivated by a feature of the PFC
model that it incorporates the properties of both liquid and solid phases, given that
the PFC free energy terms are connected to the direct correlation functions of the liquid
phase.\cite{prb64107,pre21605} Although for liquids the PFC amplitude $A_{\mathbf{K}}=0$,
indicating the limited capacity in describing the elastic behavior, the elastic response
of a liquid system can be deducted from the variation of average density $\bar{\psi}$
(i.e., zeroth mode) or system volume $V$. In addition, the liquid-state analysis is adopted
here to provide an insufficient but necessary test. We will test whether the proposed
formulation could reproduce some well recognized properties of liquids or isotropic fluids
(particularly zero shear modulus and a Poisson's ratio of 1/2). Although passing the test
does not guarantee the validity of the formulation (which needs a combination with the study
of crystalline state described in the next section), failing the test definitely indicates
that the formulation is improper. For this purpose the procedure given below is general
and not limited to the PFC model.

For liquids, the strain influences $F$ via the deformation of volume $V$. Up to second
order we have
\begin{equation}
  F = F_0+ \left. \frac{d F}{d V} \right |_{V_0} dV
  + \left. \frac{1}{2}\frac{d^2 F}{d V^2} \right |_{V_0} (dV)^2,
\label{Eq4-1}
\end{equation}
where $dV=V-V_0$. Here the free energy is expanded with respect to the undeformed 
(unstrained) state with $V=V_0$. This expansion form is used for the calculation of
elastic constants which requires the evaluation at the limit of zero strains.
Under any strain imposed on the system, no shear stress will be generated
in liquids, i.e., 
\begin{equation}
C_{ijkl}=0, \qquad {\rm for~} i\neq j {\rm ~or~} k\neq l.
\label{Eq4-2m1}
\end{equation}
Liquids are isotropic, and hence it is required that 
\begin{eqnarray}
C_{1111}=C_{2222}=C_{3333}\equiv \bar{C}_{1111}, \nonumber \\
C_{1122}=C_{1133}=C_{2233}\equiv \bar{C}_{1122}.
\label{Eq4-2}
\end{eqnarray}
In addition, Poisson's ratio of liquid is equal to 1/2, i.e.,
\begin{equation}
\nu = \frac{\bar{C}_{1122}} {\bar{C}_{1111}+\bar{C}_{1122}}=\frac{1}{2}.
\label{Eq4-3}
\end{equation}
In the following Eqs.~(\ref{Eq4-2m1})--(\ref{Eq4-3}) are used as criteria to justify
the validity of various definitions of elastic constants.

For an infinitesimal strain $\boldsymbol{\varepsilon}$, $V$ can be expanded according
to Eq.~(\ref{Eq4-4}). Here $\varepsilon_{ij}$ are treated as nine independent variables
in calculations, although with the symmetry of $\varepsilon_{ij}=\varepsilon_{ji}$. 
Substituting Eq.~(\ref{Eq4-4}) into Eq.~(\ref{Eq4-1}) yields
\begin{eqnarray}
F(\boldsymbol{\varepsilon}) &=& F_0 + \left. \frac{d F}{d V} \right |_{V_0} V_0
\left( \varepsilon_{11}+\varepsilon_{22}+\varepsilon_{33}
+ \varepsilon_{11}\varepsilon_{22} \right. \nonumber\\
&& \left.+\varepsilon_{11}\varepsilon_{33}
+\varepsilon_{22}\varepsilon_{33}-\varepsilon_{12}\varepsilon_{21}
-\varepsilon_{13}\varepsilon_{31}-\varepsilon_{23}\varepsilon_{32} \right) \nonumber \\
&& + \frac{1}{2} \left. \frac{d^2 F}{d V^2} \right |_{V_0} V_0^2
\left( \varepsilon_{11}^2+\varepsilon_{22}^2+\varepsilon_{33}^2 \right. \nonumber\\
&& \left. +2\varepsilon_{11}\varepsilon_{22}+2\varepsilon_{11}\varepsilon_{33}
+2\varepsilon_{22}\varepsilon_{33} \right)+\mathcal{O}(\boldsymbol{\varepsilon}^3) \nonumber\\
&=& F_0 + \left. \frac{d F}{d V} \right |_{V_0} V_0
\left[ \varepsilon_{ii} + \frac{1}{2} \left( \varepsilon_{ii}\varepsilon_{jj}
  -\varepsilon_{ij}\varepsilon_{ji} \right) \right] \nonumber\\
&& + \frac{1}{2} \left. \frac{d^2 F}{d V^2} \right |_{V_0} V_0^2 \varepsilon_{ii}\varepsilon_{jj}
+\mathcal{O}(\boldsymbol{\varepsilon}^3),
\label{Eq4-5}
\end{eqnarray}
which satisfies the condition of strain invariance under any orthogonal transformation.
Such an invariant condition is obeyed at any orders of $F$ expansion, given that $F$ is
expanded as a power series of $dV=V-V_0$ and volume $V$ is invariant [see Eq.~(\ref{Eq4-4})].

If defining the elastic constants as
\begin{equation}
  C_{ijkl}^{(\varepsilon)} = \frac{1}{V_0} \left. \frac{\partial^2 F}
  {\partial \varepsilon_{ij} \partial \varepsilon_{kl}} \right |_{\boldsymbol{\varepsilon}=0},
\label{Eq4-6}
\end{equation}
where the superscript ``${(\varepsilon)}$'' is used to distinguish from the definition
in Eq.~(\ref{Eq1-26a}), we have
\begin{align}
  &C_{1111}^{(\varepsilon)} = C_{2222}^{(\varepsilon)} = C_{3333}^{(\varepsilon)}
  \equiv \bar{C}_{1111}^{(\varepsilon)} = V_0 \left. \frac{d^2 F}{d V^2} \right |_{V_0} , \nonumber \\
  &C_{1122}^{(\varepsilon)}=C_{1133}^{(\varepsilon)}=C_{2233}^{(\varepsilon)}
  \equiv \bar{C}_{1122}^{(\varepsilon)} =\left. \frac{d F}{d V} \right |_{V_0}
  + V_0 \left. \frac{d^2 F}{d V^2} \right |_{V_0} , \nonumber \\
  &C_{1221}^{(\varepsilon)}=C_{1331}^{(\varepsilon)}=C_{2332}^{(\varepsilon)}
  \equiv \bar{C}_{1221}^{(\varepsilon)} = - \frac{1}{2} \left. \frac{d F}{d V} \right |_{V_0}, \nonumber \\
  &C_{\rm others}^{(\varepsilon)}=0.
\label{Eq4-7}
\end{align}
Therefore, the Poisson's ratio is given by
\begin{equation}
  \nu^{(\varepsilon)}=\frac{1}{2} \left ( 1 + \frac{ {d F}/{d V} }
           {{d F}/{d V} + 2V_0 {d^2 F}/{d V^2}} \right )_{V=V_0},
\label{Eq4-8}
\end{equation}
which generally is not equal to 1/2 when ${d F}/{d V} |_{V_0} \neq 0$ (e.g., in PFC models
giving nonzero system pressure). $\bar{C}_{1221}^{(\varepsilon)}$ obtained from
Eq.~(\ref{Eq4-7}) is not zero either. All these indicate that the definition of
Eq.~(\ref{Eq4-6}) for elastic constants is improper.

For the finite strain $\mathbf{E}$ given in Eq.~(\ref{Strain-4}), or equivalently
\begin{eqnarray}
  E_{ij}&=&\varepsilon_{ij}+\frac{1}{2} \left ( \varepsilon_{ki}+\omega_{ki} \right )
  \left ( \varepsilon_{kj}+\omega_{kj} \right) \nonumber\\
  &=& \varepsilon_{ij}+\frac{1}{2}\varepsilon_{ki}\varepsilon_{kj}
  + \mathcal{O}({\bm \omega}),
\label{Eq4-9}
\end{eqnarray}
given the absence of pure rotation effect in the system energy, 
the volume is expanded as
\begin{eqnarray}
  V &=& V_0 \left[ 1 + E_{11}+E_{22}+E_{33} +E_{11}E_{22}+E_{11}E_{33}
    \right. \nonumber \\
  && +E_{22}E_{33} -E_{12}E_{21}-E_{13}E_{31}-E_{23}E_{32} \nonumber\\
  && - \frac{1}{2} \left( E_{11}^2+E_{22}^2+E_{33}^2+E_{12}^2+E_{21}^2  \right. \nonumber\\
  && \left. \left. +E_{13}^2+E_{31}^2+E_{23}^2+E_{32}^2 \right) \right]
  +\mathcal{O}(\mathbf{E}^3) \nonumber\\
  &=& V_0 \left ( 1 + E_{ii} + \frac{1}{2} E_{ii} E_{jj} - E_{ij} E_{ji} \right )
  +\mathcal{O}(\mathbf{E}^3).
\label{Eq4-10}
\end{eqnarray}
The free energy becomes
\begin{eqnarray}
  F(\mathbf{E}) &=& F_0+ \left. \frac{d F}{d V} \right |_{V_0} V_0
  \left[ E_{11}+E_{22}+E_{33}+E_{11}E_{22} \right. \nonumber \\
  &+& E_{11}E_{33}+E_{22}E_{33} -E_{12}E_{21}-E_{13}E_{31}-E_{23}E_{32} \nonumber \\
  &-& \frac{1}{2} \left(E_{11}^2+E_{22}^2+E_{33}^2+E_{12}^2+E_{21}^2+E_{13}^2+E_{31}^2
    \right. \nonumber \\
  &+& \left. \left. E_{23}^2+E_{32}^2 \right) \right]
  +\frac{1}{2} \left. \frac{d^2 F}{d V^2} \right |_{V_0} V_0^2
  \left( E_{11}^2+E_{22}^2+E_{33}^2 \right. \nonumber\\
  &+& \left. 2E_{11}E_{22}+2E_{11}E_{33}+2E_{22}E_{33}\right) +\mathcal{O}(\mathbf{E}^3) \nonumber\\
  &=& F_0+ \left. \frac{d F}{d V} \right |_{V_0} V_0
  \left ( E_{ii} + \frac{1}{2} E_{ii} E_{jj} - E_{ij} E_{ji} \right ) \nonumber\\
  && + \frac{1}{2} \left. \frac{d^2 F}{d V^2} \right |_{V_0} V_0^2 E_{ii} E_{jj}
  +\mathcal{O}(\mathbf{E}^3),
\label{Eq4-11}
\end{eqnarray}
satisfying the strain invariant condition. 
If we define the elastic constants as
\begin{equation}
  C_{ijkl}^{(E)} = \frac{1}{V_0} \left.
  \frac{\partial^2 F} {\partial E_{ij} \partial E_{kl}} \right |_{\mathbf{E}=0},
\label{Eq4-12}
\end{equation}
where the superscript ``$(E)$'' is used to distinguish from the definitions in
Eqs.~(\ref{Eq1-26a}) and (\ref{Eq4-6}), the results are
\begin{align}
  &C_{1111}^{(E)} = C_{2222}^{(E)} = C_{3333}^{(E)} \equiv \bar{C}_{1111}^{(E)}
  = V_0 \left. \frac{d^2 F}{d V^2} \right |_{V_0} - \left. \frac{d F}{d V} \right |_{V_0}, \nonumber \\
  &C_{1122}^{(E)}=C_{1133}^{(E)}=C_{2233}^{(E)} \equiv \bar{C}_{1122}^{(E)}
  = V_0 \left. \frac{d^2 F}{d V^2} \right |_{V_0} + \left. \frac{d F}{d V} \right |_{V_0}, \nonumber \\
  &C_{1221}^{(E)}=C_{1331}^{(E)}=C_{2332}^{(E)} \equiv \bar{C}_{1221}^{(E)} =
  - \left. \frac{dF}{dV} \right |_{V_0}, \nonumber\\
  & C_{\rm others}^{(E)}=0.
\label{Eq4-13}
\end{align}
The corresponding Poisson's ratio is
\begin{equation}
\nu^{(E)}=\frac{1}{2} \left ( 1 + \frac{{d F}/{d V}}{V_0 {d^2 F}/{d V^2}} \right )_{V=V_0},
\label{Eq4-14}
\end{equation}
which generally would not give the value of 1/2 at nonzero ${d F}/{d V}|_{V_0}$, and
$\bar{C}_{1221}^{(E)} \neq 0$. Thus the definition of Eq.~(\ref{Eq4-12}) is also improper.
It is noted that these finite strain results of free energy and elastic constants
[i.e., Eqs.~(\ref{Eq4-11})--(\ref{Eq4-14})] can be reduced to those of
Eqs.~(\ref{Eq4-5})--(\ref{Eq4-8}) at the limit of infinitesimal strain, by simply
substituting Eq.~(\ref{Eq4-9}) for the expression of $E_{ij}$ into Eq.~(\ref{Eq4-11})
and keeping up to second order of $\boldsymbol{\varepsilon}$ at small strains.

For the finite strain tensor $\boldsymbol{\xi}$ defined in Eq.~(\ref{Eq1-20a}), 
the volume is expanded to be
\begin{equation}
  V =  V_0 \left( 1+\xi_{11}+\xi_{22}+\xi_{33} \right)+\mathcal{O}(\boldsymbol{\xi}^3),
\label{Eq4-16}
\end{equation}
without the second-order terms of $\boldsymbol{\xi}$. The free energy expansion is written as
\begin{eqnarray}
  F(\boldsymbol{\xi}) &=& F_0+ \left. \frac{d F}{d V} \right |_{V_0} V_0
  \left( \xi_{11}+\xi_{22}+\xi_{33} \right) \nonumber \\ 
  && + \frac{1}{2} \left. \frac{d^2 F}{d V^2} \right |_{V_0} V_0^2
  \left( \xi_{11}^2+\xi_{22}^2+\xi_{33}^2 \right. \nonumber\\
  && \left. +2\xi_{11}\xi_{22}+2\xi_{11}\xi_{33}+2\xi_{22}\xi_{33} \right)
  +\mathcal{O}(\boldsymbol{\xi}^3) \label{Eq4-17}\\
  &=& F_0+ \left. \frac{d F}{d V} \right |_{V_0} V_0 \xi_{ii}
  + \frac{1}{2} \left. \frac{d^2 F}{d V^2} \right |_{V_0} V_0^2 \xi_{ii} \xi_{jj}
  +\mathcal{O}(\boldsymbol{\xi}^3). \nonumber
\end{eqnarray}
If defining the elastic constants as the second-order derivatives of $F$ with respective
to $\boldsymbol{\xi}$ as in Eq.~(\ref{Eq1-26a}), we obtain
\begin{align}
  &C_{1111} = C_{2222} = C_{3333} \equiv \bar{C}_{1111}
  = V_0 \left. \frac{d^2 F}{d V^2} \right |_{V_0}, \nonumber \\
  &C_{1122}=C_{1133}=C_{2233} \equiv \bar{C}_{1122}
  = V_0 \left. \frac{d^2 F}{d V^2} \right |_{V_0}, \nonumber \\
  &C_{\rm others}=0,
\label{Eq4-19}
\end{align}
which gives a Poisson's ratio of $\nu=1/2$, satisfying the requirement of
Eq.~(\ref{Eq4-3}). In addition, $C_{ijkl}=0$ when $i\neq j$ or $k\neq l$, consistent
with Eq.~(\ref{Eq4-2m1}). The same results can be obtained from Eq.~(\ref{Eq1-1c})
with $G=G(V(\boldsymbol{\varepsilon}))$ for liquids.
Thus, only the definition of Eq.~(\ref{Eq1-26a}) [or equivalently Eq.~(\ref{Eq1-1c})]
passes the test of Eqs.~(\ref{Eq4-2m1})--(\ref{Eq4-3}) for liquids when
${d F}/{d V}|_{V_0} \neq 0$ as in PFC models with nonzero intrinsic pressure.

\section{Numerical results for crystal}
\label{sec:results}

\subsection{Anisotropic amplitude variations under strain}

Here we consider the PFC model parameterized for bcc Fe which has been extensively studied. 
The parameters are adopted from the work of Wu {\it et al.} \cite{prb184107} for solid-liquid
coexistence of Fe: $b=-2.136$ eV~\AA$^3$, $\lambda=0.291$ eV~\AA$^7$, $q_0=2.985$~\AA$^{-1}$,
and $g=9.705$ eV~\AA$^9$ in the PFC free energy functional Eq.~(\ref{PFC-1}); also the
average atomic density is $\bar{\rho}=0.0765$~\AA$^{-3}$.
In the dimensionless form, we have $\epsilon=0.0923$, and the corresponding average
rescaled density at solid-liquid coexistence is $\bar{\psi}=-0.201$. \cite{prb14103}  
Combining Eqs.~(\ref{PFC-2}) and (\ref{PFC-12}) and the rescaling of $\psi$ yields
\begin{equation}
\bar{\psi}= \sqrt{\frac{g}{\lambda q_0^4}} \left(\bar{\rho}-\rho_0\right)
=\sqrt{\frac{g}{\lambda q_0^4}} \bar{\rho}-\tilde{\rho}_0,
\label{Res1}
\end{equation}
based on which we have $\tilde{\rho}_0=0.251$. Elastic properties of this PFC system
have been evaluated numerically, with the calculation procedure described in
Appendix~\ref{sec:appen_NDetails}.

\begin{figure*}
\includegraphics[width=0.7\textwidth]{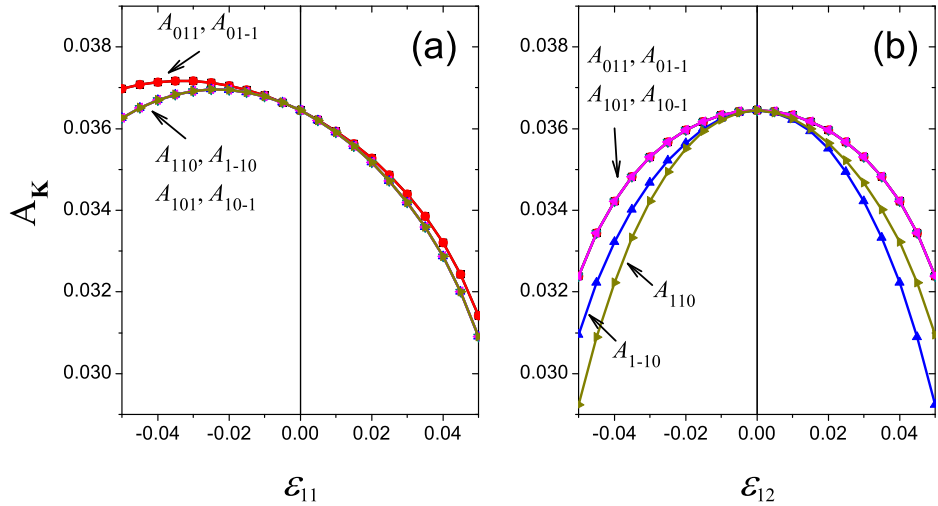}
\caption{Amplitudes $A_{\mathbf{K}}$ as a function of
  (a) uniaxial strain $\varepsilon_{11}$ and (b) shear strain $\varepsilon_{12}$,
  for bcc Fe with $\epsilon=0.0923$, $\bar{\psi}=-0.201$, and $\tilde{\rho}_0=0.251$.
\label{fig2}}
\end{figure*}

For an unstrained bcc structure, due to the crystal symmetry the values of the first-mode
amplitudes $A_{\mathbf K}$ are equal to each other as given in Eq.~(\ref{PFC-17}). However,
when a strain is applied, leading to anisotropic deformation of the lattice, the degeneracy
of $A_{\mathbf K}$ is broken and the six first-mode amplitudes should be evaluated independently. 
The numerical variations of $A_{\mathbf K}$ for a bcc Fe under a uniaxial and a shear strain
are shown in Fig.~\ref{fig2}. Two amplitudes $A_{011}$ and $A_{01\bar{1}}$, for which the
wave vectors $\mathbf{K}$ are perpendicular to the applied direction of the uniaxial
strain, become larger than the other four amplitudes [see Fig.~\ref{fig2}(a)]. 
This is consistent with the observation of H\"{u}ter {\it et al.}.\cite{prb214105} 
In addition, the slope of the $A_{\mathbf K}$ vs $\varepsilon_{11}$ curves is nonzero at
$\varepsilon_{11}=0$, indicating that the variations of $A_{\mathbf K}$ subjected to infinitesimal
uniaxial elastic deformation is not negligible. Under a shear strain, values of $A_{\mathbf K}$
are split into three groups: $A_{110}$, $A_{1\bar{1}0}$, and the other four, as shown in
Fig.~\ref{fig2}(b).

\subsection{Influence of various variation schemes of average atomic density under deformation}

\begin{figure}
\includegraphics[width=0.45\textwidth]{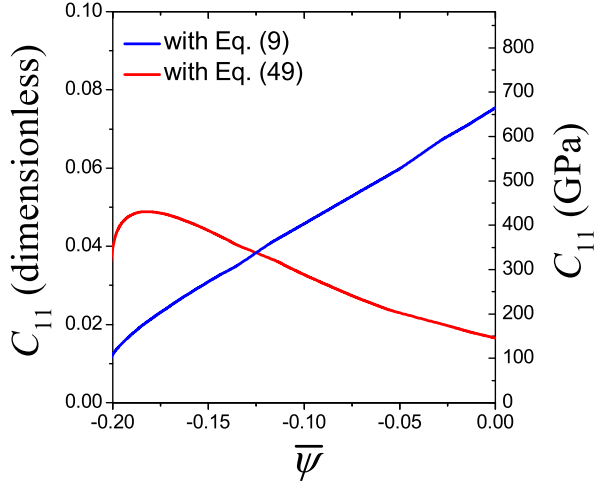}
\caption{Elastic constant $C_{11}$ as a function of $\bar{\psi}$, for a bcc phase with
  $\epsilon=0.0923$ and $\tilde{\rho}_0=0.251$. The variation of $\bar{\psi}$ under strain
  follows Eq.~(\ref{PFC-11}) or improper Eq.~(\ref{Res2}), giving results as blue or red
  curves, respectively. $C_{ij}$ is calculated via Eq.~(\ref{Eq1-26a}), and is measured
  both in a dimensionless unit (left axis) and a physical unit (right axis).
\label{fig3}}
\end{figure}

In PFC models, various definitions of atomic density field ($\rho$, $\phi$, $n$, or $\psi$
as described above) can be used with a very similar form of free energy functional. 
However, their variations in response to strain or external deformation are different as
shown in Eqs.~(\ref{PFC-8})--(\ref{PFC-11}). This causes some confusion or discrepancies
in previous studies. For example, Pisutha-Arnond {\it et al.}\cite{prb14103} have
pointed out the importance of volume and density variations in elastic response, 
but applied the variation scheme of 
\begin{equation}
\bar{\psi}_{\rm strained}=\frac{V_{\rm unstrained}}{V_{\rm strained}} \bar{\psi}_{\rm unstrained},
\label{Res2}
\end{equation}
instead of Eq.~(\ref{PFC-11}) for $\bar{\psi}$, which implies that in Ref.~\onlinecite{prb14103}
the variation of $\bar{\psi}$ under deformation was interpreted as that of $\bar{\rho}$.
(Note that $\rho$ is atomic density and is always positive, while $\phi$, $n$ and $\psi$ are
density differences, not necessarily of positive values.)
The variation scheme of $\bar{\psi}$ has important influence on the resulting elastic
constants. Some results of our numerical calculations based on the elastic constant
definition of Eq.~(\ref{Eq1-26a}) are presented in Fig.~\ref{fig3},
where Voigt notation has been used, i.e., $C_{11}=C_{1111}$, $C_{22}=C_{2222}$, $C_{33}=C_{3333}$,
$C_{12}=C_{1122}$, and $C_{44}=C_{2323}$. If using the scheme of Eq.~(\ref{Res2}), the calculated
value of $C_{11}$ first increases and then decreases with increasing $\bar{\psi}$ (red line in
Fig.~\ref{fig3}). A similar trend has been observed in the work of Pisutha-Arnond {\it et al.},
\cite{prb14103} although Eq.~(\ref{Eq4-12}) was used there in calculating elastic constants.
When the proper scheme in Eq.~(\ref{PFC-11}) is adopted, $C_{11}$ monotonously increases
with $\bar{\psi}$ (blue line in Fig.~\ref{fig3}), as usually expected. The obtained $C_{11}$
value for bcc Fe at $\bar{\psi}=-0.201$ is 109 GPa, close to the MD result of 128 GPa.
\cite{pre61601} This value is much smaller than that obtained with Eq.~(\ref{Res2})
(and that in Ref.~\onlinecite{prb14103}), suggesting that the overestimation of $C_{11}$
in the previous study is caused more by the used algorithm for elastic response,
than the inaccuracy in PFC fitting parameters.

\subsection{Role of the linear term in free energy functional}

\begin{figure*}
\includegraphics[width=\textwidth]{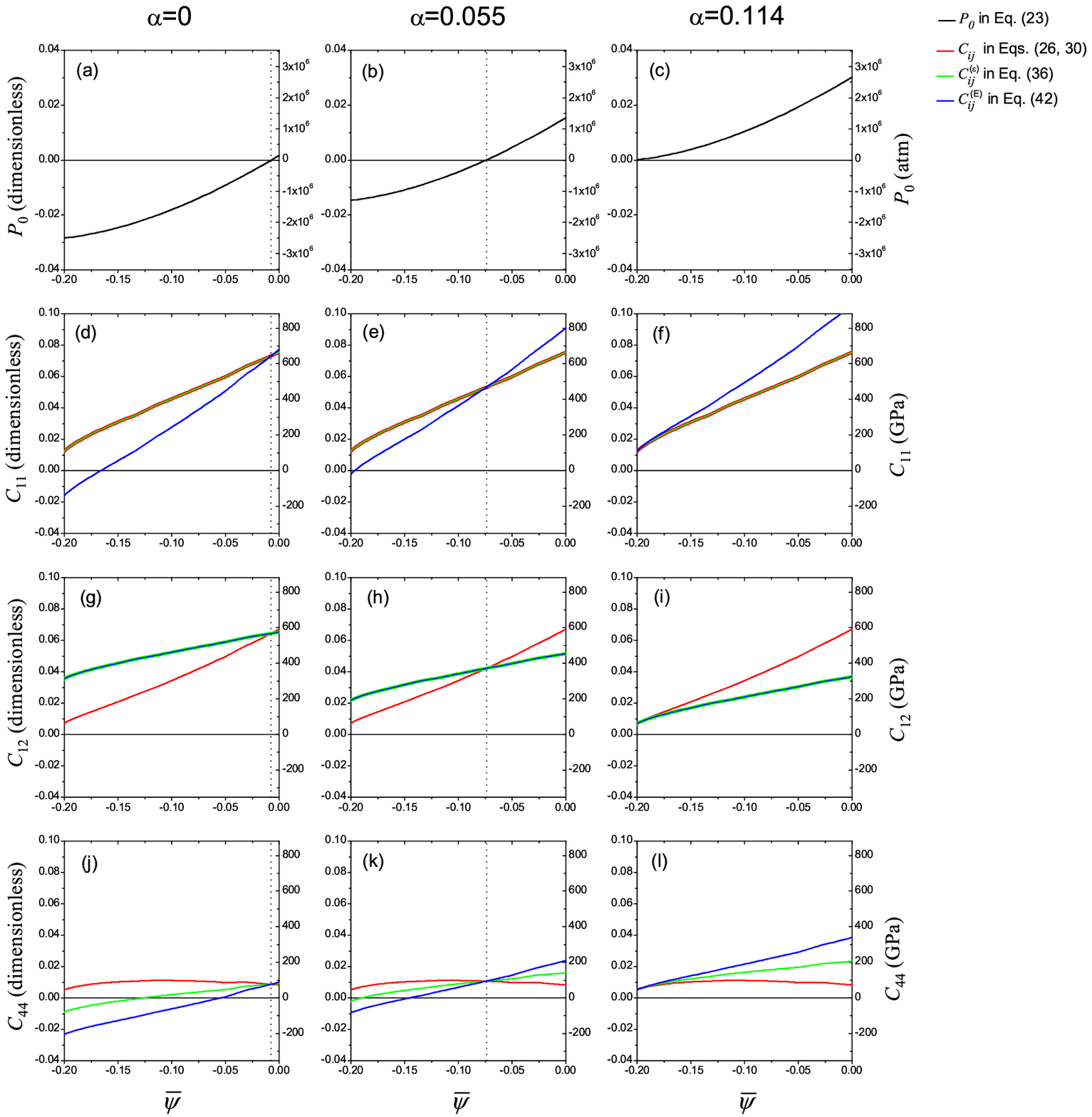}%
\caption{The pressure and elastic constants as functions of $\bar{\psi}$, for a bcc phase with
  $\epsilon=0.0923$ and $\tilde{\rho}_0=0.251$. Equation~(\ref{PFC-11}) is used for the variation
  of $\bar{\psi}$ under elastic deformations. Results of three different $\alpha$ values for
  the linear term in the free energy functional Eq.~(\ref{PFC-4}) are shown, with $\alpha=0$
  (left panels), $\alpha=0.055$ (middle panels), and $\alpha=0.114$ (right panels). 
  Elastic constants are calculated via Eq.~(\ref{Eq1-1c}) or (\ref{Eq1-26a}) for $C_{ij}$
  (red lines), Eq.~(\ref{Eq4-6}) for $C^{(\varepsilon)}_{ij}$ (green lines), or Eq.~(\ref{Eq4-12})
  for $C^{(E)}_{ij}$ (blue lines). The $\bar{\psi}$ value at which $P_0=0$ is indicated by
  vertical dashed line.
\label{fig4}}
\end{figure*}

In previous studies, the linear term in the PFC free energy functional [$a\phi$ in Eq.~(\ref{PFC-1})
and $\alpha\psi$ in Eq.~(\ref{PFC-4})] usually was not included since it gives a constant
($aV\bar{\phi}$ or $\alpha V\bar{\psi}$) after integration. However, when we consider the
pressure $P_0$ and the elastic constant $C_{ij}$, both $V$ and $\bar{\psi}$ (or $\bar{\phi}$) 
change with elastic deformation (strain). Thus the linear term is important for $P_0$ and
$C_{ij}$ calculations and cannot be neglected. We have conducted numerical calculations
for a bcc phase (using PFC parameters for Fe as
described above), based on Eq.~(\ref{PFC-11}) for the variation of $\bar{\psi}$ with
volume $V$, Eq.~(\ref{Eq1-1b}) for $P_0$, and three different elastic constant formulae of
Eq.~(\ref{Eq1-1c}) or (\ref{Eq1-26a}) for $C_{ij}$, Eq.~(\ref{Eq4-6}) for $C_{ij}^{(\varepsilon)}$,
and Eq.~(\ref{Eq4-12}) for $C_{ij}^{(E)}$. Detailed results are given in Fig.~\ref{fig4}
for different values of $\alpha$.

When $\alpha=0$, $P_0$ of bcc Fe calculated from the PFC model with $\bar{\psi}=-0.201$
can be as high as $-2.5\times 10^6$ atm [Fig.~\ref{fig4}(a)]. This indicates that neglecting
the linear term in the PFC free energy functional would lead to an unrealistic value of
pressure. The zero point of $P_0$ locates very close to $\bar{\psi}=0$ at $\alpha=0$, 
and it moves to a smaller value of $\bar{\psi}$ with increasing $\alpha$, as shown in
Fig.~\ref{fig4}(a)--(c). When $\alpha=0.114$, at $\bar{\psi}=-0.201$ which is the
parameter fitted for bcc Fe, we get $P_0=0$, a thermodynamic state used in MD simulations.
\cite{pm3977} In addition, $P_0$ increases with increasing $\bar{\psi}$, consistent
with the expectation of larger pressure at higher atomic density. On the other hand, if
the improper Eq.~(\ref{Res2}) is used instead of Eq.~(\ref{PFC-11}) for the $\bar{\psi}$
varying scheme, $P_0$ abnormally decreases with increasing $\bar{\psi}$ (data not shown here),
which was also found in Ref.~\onlinecite{prb14103}.

Figures \ref{fig4}(d)--(l) gives results of different sets of elastic constants that are
formulated via Eq.~(\ref{Eq1-1c}), (\ref{Eq4-6}), or (\ref{Eq4-12}). Only $C_{ij}$ defined
in Eq.~(\ref{Eq1-1c}), or equivalently in Eq.~(\ref{Eq1-26a}), are independent of the
choice of $\alpha$ [i.e., of the value of pre-existing pressure $P_0$ in the model; see red
lines in Fig.~\ref{fig4}(d)--(l)]. The other two sets of elastic constants $C_{ij}^{(\varepsilon)}$
and $C_{ij}^{(E)}$ defined in Eqs.~(\ref{Eq4-6}) and (\ref{Eq4-12}) both change with the
$\alpha$ value used [green and blue lines in Fig.~\ref{fig4}(d)--(l)], and more seriously,
they may even become negative [see Fig.~\ref{fig4}(d), (e), (j), and (k)] although the
corresponding undeformed states are stable. The difference among these different sets of
elastic constants arises from nonzero $P_0$, which could be of huge value as shown in
Fig.~\ref{fig4}(a)--(c). From Eqs.~(\ref{Eq1-1c}), (\ref{Eq4-6}), and (\ref{Eq4-12}),
for a stable state under pressure $P_0$, i.e., $\sigma_{ij}^{(0)}=-P_0\delta_{ij}$, 
we have
\begin{eqnarray}
  &&C_{ijkl}^{(\varepsilon)}=C_{ijkl}-P_0\left(\delta_{ij}\delta_{kl}
  -\delta_{ik}\delta_{jl}/2-\delta_{il}\delta_{jk}/2 \right), \label{Res3} \\
  &&C_{ijkl}^{(E)}=C_{ijkl}-P_0\left(\delta_{ij}\delta_{kl}-\delta_{ik}\delta_{jl}
  -\delta_{il}\delta_{jk}\right). \label{Res4}
\end{eqnarray}
Note that Eq.~(\ref{Res4}) is the same as Eq.~(2.56) in Ref.~\onlinecite{pr776}, indicating
that in the special case of pre-existing isotropic stress or pressure $P_0$, $C_{ijkl}$
introduced here is equivalent to the stress-strain elastic coefficient $B_{ijkl}$ (the
generalized Birch's coefficient \cite{pr809,pr776}).

When $P_0=0$, $C_{ijkl}$, $C_{ijkl}^{(\varepsilon)}$, and $C_{ijkl}^{(E)}$ are identical, as can
be seen in Fig.~\ref{fig4}(d)--(l) where different curves of $C_{ij}$ overlap at a $\bar{\psi}$
value corresponding to $P_0=0$ (indicated by vertical dashed line). Therefore, when the linear
term of PFC free energy functional is introduced to account for $P_0=0$ or $P_0\approx 0$ 
emulating normal experimental conditions, different formulations of elastic constants are
consistent with each other and all can be adopted. In other cases with nonzero pre-existing
system pressure $P_0$ (particularly when the linear term is neglected in the model), only the
formulation of elastic constants given in Eqs.~(\ref{Eq1-1c}) and (\ref{Eq1-26a}), in terms
of Gibbs free energy or the finite strain tensor ${\bm \xi}$, gives proper results comparable
to those of real systems and should be used.

\subsection{Poisson's ratio is not 1/3}

\begin{figure}
\includegraphics[width =0.45\textwidth]{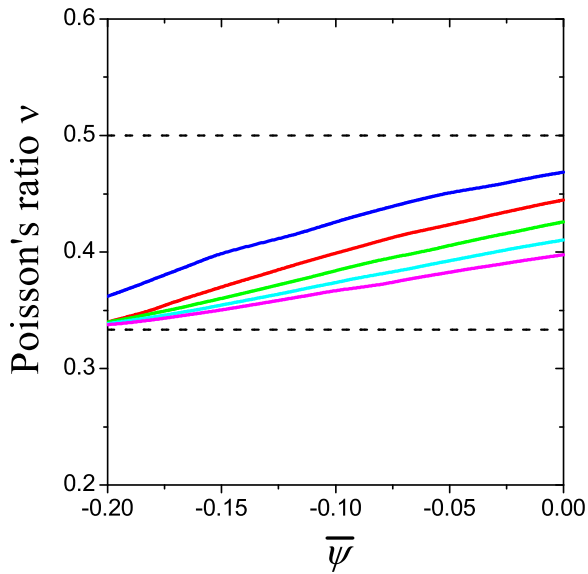}%
\caption{Poisson's ratio $\nu$ as a function of $\bar{\psi}$ for a bcc phase with 
$\tilde{\rho}_0=0.251$ and $\epsilon=0.1$, 0.2, 0.3, 0.4, and 0.5 (from top to bottom). 
The dashed lines correspond to $\nu=1/3$ and $\nu=1/2$.}
\label{fig5}
\end{figure}

Different from Eq.~(\ref{Strain-7}) 
where $\mathbf{K}$, $A_{\mathbf{K}}$, $\bar{\psi}$, and $V$ all vary with the applied
strain, the previous studies incorporated only part of the variations. As described
in Sec.~\ref{sec:strain}, most studies considered only the variation of $\mathbf{K}$,
\cite{prl245701,prl35501,prl205502,pre51605,prb125408,pre61601,pre11602}
for which the strained-state free energy is of the form
\begin{eqnarray}
  && F_{\rm strained}= \nonumber \\
  && \mathcal{F}\left( \left\{A_{\mathbf{K}}^{(0)}\right\},
  \left\{\mathbf{K}^{\rm (strained)}\right\}, \bar{\psi}_{\rm unstrained}, V_{\rm unstrained} \right),
\label{Res6}
\end{eqnarray}
where $A_{\mathbf{K}}^{(0)}$ represents the equilibrium amplitude of the unstrained
state. Only two recent works examined the additional factor of anisotropic variation of
$A_{\mathbf{K}}$ (with $\bar{\psi}$ and $V$ unchanged),\cite{prb214105} with
\begin{eqnarray}
  && F_{\rm strained}= \nonumber \\
  && \min_{\left\{A_{\mathbf{K}}\right\}} \mathcal{F}\left( \left\{A_{\mathbf{K}}\right\},
  \left\{\mathbf{K}^{\rm (strained)}\right\}, \bar{\psi}_{\rm unstrained}, V_{\rm unstrained} \right),
\label{Res5}
\end{eqnarray}
or the deformation dependence of $\bar{\psi}$ and $V$ (with $A_{\mathbf{K}}$ unchanged),
\cite{prb14103} with
\begin{eqnarray}
  && F_{\rm strained}= \nonumber \\
  && \mathcal{F}\left( \left\{A_{\mathbf{K}}^{(0)}\right\},
  \left\{\mathbf{K}^{\rm (strained)}\right\}, \bar{\psi}_{\rm strained}, V_{\rm strained} \right).
  \label{Res5_2}
\end{eqnarray}
Actually Eq.~(\ref{Res6}) and Eq.~(\ref{Res5}) are equivalent in determining elastic
constants for the PFC free energy functional Eq.~(\ref{PFC-4}), since the resulting
first-order variation of $A_{\mathbf{K}}$ is equal to zero (see Appendix \ref{sec:appen_AK}
for a general proof) and the contribution of $A_{\mathbf{K}}$ variation in the change
of free energy is beyond the second order of strain. \cite{pre61601} From this approach the
Poisson's ratio $\nu$ calculated in the one-mode approximation is always equal to 1/3, for
different PFC model parameters and average atomic density. \cite{pre51605,pre61601,pre11602} 
However, when the full variation of Eq.~(\ref{Strain-7}) is adopted, $\nu$ is no longer
restricted to 1/3. Instead, as shown in Fig.~\ref{fig5}, $\nu$ varies within the range
between 1/3 and 1/2, and increases with greater $\bar{\psi}$ and decreases with larger
$\epsilon$ value (lower temperature).

The values of $C_{ij}$ and $\nu$ for bcc Fe determined by various algorithms are listed
in Table \ref{Table1}. Among them, the method developed in this study produces overall
more reasonable results of elastic constants that are closest to the quantities obtained
by the MD simulation. It is noted that the gradient terms of the PFC model used here are
based only on the two-point direct correlation and isotropic pair interaction of the system,
\cite{prb64107,pre21605,pre11602} 
which may cause the underestimation of elastic constants. It is expected that with the
incorporation of three- and four-point correlations, \cite{WangZL} the results of $C_{ij}$
and $\nu$ in the PFC model would be improved to better match the real materials.

\begin{table}
  \caption{Elastic constants $C_{ij}$ (in units of GPa) and Poisson's ratio $\nu$ of bcc Fe
    evaluated from various algorithms. Parameters $\epsilon=0.0923$ and $\tilde{\rho}_0=0.251$
    are used in the PFC model. PFC-WAK \cite{pre61601} only considered the change of
    $\mathbf{K}$ [i.e., Eq.~(\ref{Res6}) for free energy in the strained state],
    while PFC-PCET\cite{prb14103} incorporated the variations of $\mathbf{K}$, $\bar{\psi}$,
    and $V$ but neglected that of $A_{\mathbf{K}}$ [i.e., Eq.~(\ref{Res5_2})] and used
    Eqs.~(\ref{Eq4-12}) and (\ref{Res2}). In this work, the calculations are based on
    Eqs.~(\ref{PFC-11}), (\ref{Strain-7}), and (\ref{Eq1-26a}).}
\begin{ruledtabular}
\begin{tabular}{c|cccc}
 & $C_{11}$  & $C_{12}$  & $C_{44}$   & $\nu$  \\
\hline 
MD\cite{pre61601} & 128.0 & 103.4 & 63.9 & 0.446 \\
PFC-WAK\cite{pre61601} & 90.0 & 45.0 & 45.0 & 0.333 \\
PFC-PCET\cite{prb14103} & 542.0 & 128.1 & 229.4 & 0.191 \\
This work & 109.3 & 62.7 & 46.0 & 0.365 \\
\end{tabular}
\end{ruledtabular}
\label{Table1}
\end{table}

\section{Remarks and Discussion}
\label{sec:discussion}

Calculation of elastic constants is important for the study of material properties via
the PFC modeling and also for the parameterization of the model. However, there are some
subtleties and inconsistencies on the determination of elastic constants in the existing
PFC efforts. In response to an imposed strain and deformation, all of $\mathbf{K}$,
$A_{\mathbf{K}}$, $\bar{\psi}$ and $V$ change accordingly. In previous studies,
\cite{prl245701,prl35501,prl205502,pre51605,prb125408,pre61601,pre11602,prb214105,prb14103} 
incomplete schemes were adopted and different algorithms were used for solids and liquids. 
For solids, the variation of $\mathbf{K}$ in elastic response has been well recognized
while the variation of $\bar{\psi}$ and $V$ was often neglected. For liquids, on the other
hand, the variation of $\bar{\psi}$ and $V$ was always considered since there is no 
nonzero $\mathbf{K}$ for liquids. With the scheme proposed in this study and the
incorporation of $\mathbf{K}$, $A_{\mathbf{K}}$, $\bar{\psi}$, and $V$ variations, both
solids and liquids can be described within a unified approach.

The treatment here for the variations of $A_{\mathbf{K}}$ and $\mathbf{K}$ under strain
is consistent with that of the amplitude expansion formulation studied before for PFC
models. Equation (\ref{PFC-13}) is also used in the amplitude expansion, but with
basic wave vectors $\mathbf{K}$ kept constant and the zeroth-mode average density
$\bar{\psi}=\psi_0(\mathbf{r},t)$ and amplitudes $A_{\mathbf{K}}=A_{\mathbf{K}}(\mathbf{r},t)$
varying with space and time, in contrast to the strain-induced change of $\mathbf{K}$
and spatial and temporal independence of $A_{\mathbf{K}}$ (assumed to be real in
the calculations here) and $\bar{\psi}$ considered in this work. However, in the
amplitude formulation $A_{\mathbf{K}}$ are complex variables, i.e.,
$A_{\mathbf{K}} = |A_{\mathbf{K}}| \exp(i \theta_{\mathbf{K}})$, and their phases
vary spatially as $\theta_{\mathbf{K}} = {\bm \delta}{\mathbf{K}} \cdot \mathbf{r}$
in the equilibrium or steady state of strained solids. \cite{prb165421,pre21605,pre11602}
This leads to a strain-dependent change of $\mathbf{K} \rightarrow \mathbf{K} +
{\bm \delta}{\mathbf{K}}$ (with ${\bm \delta}{\mathbf{K}}$ proportional to strain),
consistent with the variation determined here. Similar findings of degeneracy breaking
and anisotropy of $|A_{\mathbf{K}}|$ have also been obtained in numerical calculations
of amplitude equations. \cite{prb165421,pre21605} Since here Eq.~(\ref{Strain-5}) is
used to determine the instantaneous variation of $\mathbf{K}$ (or equivalently
$\theta_{\mathbf{K}}$) under strain and $|A_{\mathbf{K}}|$ is calculated from the subsequent
free energy minimization, our procedure is analogous to that in Ref.~\onlinecite{pre32411}
where the elastic equilibration through $\theta_{\mathbf{K}}$ is treated separately in
the amplitude formulation. For the average density $\bar{\psi}$ (or $\psi_0$), it is
noted that previous studies of amplitude expansion were conducted under the assumption
of constant system volume, leading to the conserved dynamics of $\psi_0$, while the
above analysis indicates that it would be interesting to extend the amplitude formulation
to incorporate the change of $\bar{\psi}$ with deformed volume under strain.

How to calculate elastic constants from the variation of free energy under strains is
also essential. Because the linear term in the free energy functional of PFC models was
usually ignored, the systems described are actually stressed intrinsically. For example,
the predicted pressure of bcc Fe is more than a million atms in the PFC model. \cite{prb14103}
For stressed materials, there are different types of elastic constants defined from 
thermodynamics, \cite{pr776,prb423,jasa348} including those given in Eqs.~(\ref{Eq4-6})
and (\ref{Eq4-12}) and also Eq.~(\ref{Eq1-26a}) derived here. To model normal experimental
conditions with pressure close to zero, the linear term should be included in the PFC
free energy functional with the corresponding coefficient determined by the condition of
zero pressure, so that different definitions of elastic constants would converge to yield
equivalent results. Otherwise, there is significant difference among various formulations
of elastic constants, and only that defined in Eq.~(\ref{Eq1-26a}) or (\ref{Eq1-1c}) (i.e.,
$C_{ijkl}$) is independent of the linear term and pressure and gives consistent results.

It is also important to note that this isothermal elastic constant $C_{ijkl}$ is the same
as the stress-strain $B$ coefficient \cite{pr776} $B_{ijkl}$ (a generalization of Birch's
coefficients for cubic symmetry \cite{pr809}) in the case of isotropic hydrostatic
pressure, although for more general cases of anisotropic stress they are different.
In previous studies of hydrostatically pressured materials, \cite{pr776,prb423} the
$B_{ijkl}$ coefficients are used for identifying elastic constants of the system. These
elastic coefficients are determined by the stress-strain relation \cite{pr776,pr809} but
generally do not possess complete Voigt symmetry for the cases of anisotropic stress.
In comparison, the $C_{ijkl}$ elastic constants introduced in this work are determined
through thermodynamic potential ($G$ or $F$) and always have complete Voigt symmetry.
Actually it can be proved that $C_{ijkl}$ is equivalent to the symmetric combination of
$B$ coefficients, given
\begin{eqnarray}
  C_{ijkl} &=& C_{ijkl}^{(\varepsilon)} + \frac{1}{2} \left [
    \frac{1}{2} \sigma_{ik}^{(0)}\delta_{jl} + \frac{1}{2} \sigma_{il}^{(0)}\delta_{jk}
    + \frac{1}{2} \sigma_{jk}^{(0)}\delta_{il} \right. \nonumber\\
  && \left. + \frac{1}{2} \sigma_{jl}^{(0)}\delta_{ik}
    - \sigma_{ij}^{(0)}\delta_{kl} - \sigma_{kl}^{(0)}\delta_{ij} \right ] \nonumber\\
  &=& C_{ijkl}^{(E)} + \frac{1}{2} \left [
    \sigma_{ik}^{(0)}\delta_{jl} + \sigma_{il}^{(0)}\delta_{jk}
    + \sigma_{jk}^{(0)}\delta_{il} \right. \nonumber\\
  && \left. + \sigma_{jl}^{(0)}\delta_{ik}
    - \sigma_{ij}^{(0)}\delta_{kl} - \sigma_{kl}^{(0)}\delta_{ij} \right ] \nonumber\\
  &=& \frac{1}{2} \left ( B_{ijkl} + B_{klij} \right ),
\end{eqnarray}
which can be obtained from Eqs. (\ref{Eq1-15a}) and (\ref{Eq1-20a}) and the definitions
of elastic constants. Although a similar form of symmetrized $B$ coefficient has been
used in some previous work, \cite{prb12627,pma2827} it was for the study of system
mechanical stability. Here we derive it from thermodynamic formulation (as shown in
Appendix \ref{sec:appen_elastic}) and demonstrate that it can be defined as the proper
elastic constants for the study of pre-stressed material systems.

Although the formulation constructed here that is based on Gibbs free energy is mainly
for the elastic constant calculation (which also plays an important role on the PFC
model parameterization), it can be applied to the study of system dynamics and evolution
for material simulations. In most of the existing PFC work, the dynamics of atomic
density field $\psi$ is assumed to be driven by the minimization of Helmholtz free
energy $F$, under the condition of constant temperature and constant volume. To
simulate material systems with constant pressure as in real experiments and also
set in many atomistic simulations like MD, the PFC dynamics should be driven to
minimize the Gibbs free energy $G$, which would lead to more realistic outcomes
in PFC simulations of, e.g., materials growth and structural evolution. The corresponding
detailed formulating and analysis are beyond the scope of this work and will be a subject
of our future research.

\section{Summary}
\label{sec:summary}

In summary, we have clarified the method for calculating isothermal elastic constants
of solids and liquids under pre-existing stress or pressure. When subjected to an applied
strain, the average density of the system is changed by the deformation, and the variation
formulae for various definitions of density fields ($\rho$, $n$, $\phi$, and $\psi$) are
different [Eqs.~(\ref{PFC-8})--(\ref{PFC-11})]. This leads to different results of elastic
constant calculations, indicating the importance of physical interpretation of the PFC
density field. The density amplitudes of the deformed solid also differ from the undeformed
ones, and their degeneracy is broken as a result of anisotropic lattice distortion.

Our results also show that due to the existence of high pressure in the model system
(e.g., when neglecting the linear term in the free energy functional), it is not
suitable to calculate elastic constants $C_{ijkl}$ as the second-order derivatives of the
Helmholtz free energy $F$ with respective to infinitesimal or finite strain tensor, which
would lead to unphysical results without incorporating the effect of pre-existing stress.
Instead, either a new strain tensor $\xi_{ij}$ [Eq.~(\ref{Eq1-20a})] needs to be introduced
to calculate $C_{ijkl}$ from $F$ [Eq.~(\ref{Eq1-26a})], or a Gibbs-type free energy $G$
[Eq.~(\ref{Eq1-15a})] should be used [Eq.~(\ref{Eq1-1c})]. The validity of our formulation
has been tested through an analytic calculation of elastic constants for the liquid phase,
as well as numerical calculations conducted on the PFC model parameterized for bcc Fe.
Compared to previous PFC work, the results obtained from our method are more consistent
with the data of MD simulations. Although the system studied in this work is based on the
PFC model, the approach and the elastic constant formulation developed here from thermodynamics
are generic and can be applied to the study of general stressed material systems.

\begin{acknowledgments}

Z.R.L. acknowledges support from the National Natural Science Foundation of China (Grant No.~21773002). 
Z.-F.H. acknowledges support from the National Science Foundation under Grant No.~DMR-1609625. 
The authors thank Zeren Lin for helpful discussions.

\end{acknowledgments}

\appendix
\section{Derivation of elastic constants in system under constant pre-existing stress}
\label{sec:appen_elastic}

Elastic constants $C_{ijkl}$ can be determined by examining the work to be paid
when the system is deformed from the initial state to the final strained state with
the strain tensor $\boldsymbol{\varepsilon}$. Based on the first law of thermodynamics,
in an isothermal system with constant temperature, the total work done on
the system is equal to the change of its free energy, i.e.,
\begin{equation}
W^{\rm (total)} = \Delta F.
\label{Eq1-4}
\end{equation} 
However, when there exists a pre-applied pressure or stress, the work done by it,
$W^{\rm (ext)}$, should be subtracted from $W^{\rm (total)}$ to give the actual work needed:
\begin{equation}
\Delta W = W^{\rm (total)} - W^{\rm (ext)} = \Delta F - W^{\rm (ext)}.
\label{Eq1-5}
\end{equation} 
Therefore, $\Delta F - W^{\rm (ext)}$, instead of $\Delta F$, should be used in calculating
elastic constants $C_{ijkl}$ in the presence of a pre-existing external stress.

Here we derive a formula of $W^{\rm (ext)}$ under a general, constant pre-applied stress
tensor $\boldsymbol{\sigma}^{\rm (ext)}=\{\sigma_{ij}^{\rm (ext)}\}$ when the system is deformed
homogeneously from an initial unstrained but pre-stressed state to a final state with any
specified strain $\boldsymbol{\varepsilon}$ (where the strain is measured from the
initial pre-stressed state). The pre-applied force acting on a surface element is given by
\begin{equation}
  d{\bf f}^{\rm (ext)}= \boldsymbol{\sigma}^{\rm (ext)} \cdot \hat{\bf n} d^2s
  = \boldsymbol{\sigma}^{\rm (ext)} \cdot d^2{\bf s},
\label{Eq1-6}
\end{equation} 
where $\hat{\bf n}$ represents the normal direction of the surface element $d^2{\bf s}$. 
The position vector of this element (and the related volume element) is denoted as ${\bf r}$,
while the corresponding position vector in the initial undeformed state is denoted as ${\bf R}$.
For each surface element $d^2{\bf s}$ or volume element $d^3V$ (corresponding to each
$\mathbf{R}$), the quasistatic variation of its elastic deformation can be described by
the varying of an effective strain order parameter $\tilde{\varepsilon}$, given that the
stress $\boldsymbol{\sigma}^{\rm (ext)}$ remains constant during the deformation process. Thus
\begin{equation}
{\bf r}= \left( 1+ \tilde{\varepsilon} \boldsymbol{\varepsilon} \right) \cdot \mathbf{R},
\label{Eq1-7}
\end{equation} 
where $\tilde{\varepsilon}$ represents the completion degree or state of the quasistatic
homogeneous deformation process. $\tilde{\varepsilon}=0$ corresponds to the initial
unstrained state, while $\tilde{\varepsilon}=1$ corresponds to the final
deformed state with strain $\boldsymbol{\varepsilon}$. The position displacement of
each volume element (of a given $\mathbf{R}$) during the infinitesimal process of
$\tilde{\varepsilon} \rightarrow \tilde{\varepsilon}+d\tilde{\varepsilon}$ is then
\begin{equation}
d'{\bf r} = \boldsymbol{\varepsilon} \cdot \mathbf{R} d\tilde{\varepsilon},
\label{Eq1-8}
\end{equation} 
and the work done by the pre-applied external stress on each element is $d{\bf f}^{\rm (ext)}
\cdot d'{\bf r}$. In this specific case of constant stress, the corresponding work done
should depend only on the initial and final strain states characterized by the state order
parameter $\tilde{\varepsilon}$, leading to the following result for the external work
done on the whole system:
\begin{eqnarray}
W^{\rm (ext)} &=& \int d{\bf f}^{\rm (ext)} \cdot d'{\bf r} \nonumber\\
&=& \int_0^1 d\tilde{\varepsilon}
\oiint_{\partial V} \left[ (\boldsymbol{\varepsilon} \cdot \mathbf{R}) \cdot
  \boldsymbol{\sigma}^{\rm (ext)} \right] \cdot d^2{\bf s}.
\label{Eq1-9}
\end{eqnarray} 
Using the divergence theorem, 
Eq.~(\ref{Eq1-9}) becomes 
\begin{equation}
  W^{\rm (ext)} = \int_0^1 d\tilde{\varepsilon} \iiint_V {\bm \nabla} \cdot
  \left[ (\boldsymbol{\varepsilon} \cdot \mathbf{R})
    \cdot \boldsymbol{\sigma}^{\rm (ext)} \right]  d^3V.
\label{Eq1-10}
\end{equation} 
Noting that ${\bm \nabla}$ acts on $\mathbf{r}$ while $\boldsymbol{\sigma}^{\rm (ext)}$
and $\boldsymbol{\varepsilon}$ remain constant during the homogeneous deformation,
from Eqs.~(\ref{Eq1-7}) and (\ref{Eq1-10}) we have
\begin{eqnarray}
&& W^{\rm (ext)} = \int_0^1 d\tilde{\varepsilon} \iiint_V \nabla\cdot
  \frac{1}{\tilde{\varepsilon}} \left[ \left( \mathbf{r} -\mathbf{R}\right) 
    \cdot \boldsymbol{\sigma}^{\rm (ext)} \right] d^3V \nonumber \\
&&= \int_0^1 d\tilde{\varepsilon} \iiint_V \nabla \cdot \frac{1}{\tilde{\varepsilon}}
  \left[ \left( \mathbf{r} - (1+\tilde{\varepsilon}\boldsymbol{\varepsilon})^{-1}
    \cdot \mathbf{r} \right) \cdot \boldsymbol{\sigma}^{\rm (ext)} \right] d^3V  \nonumber \\
&&= \int_0^1 d\tilde{\varepsilon} \iiint_V \frac{1}{\tilde{\varepsilon}} 
  \left[ \left( \delta_{ij} - (1+\tilde{\varepsilon}\boldsymbol{\varepsilon})^{-1}_{ij} \right) 
    \sigma^{\rm (ext)}_{ij} \right] d^3V \nonumber \\
&&= \int_0^1 \frac{1}{\tilde{\varepsilon}} \left[ 
    \left( \delta_{ij} - (1+\tilde{\varepsilon}\boldsymbol{\varepsilon})^{-1}_{ij} \right) 
    \sigma^{\rm (ext)}_{ij} \right] V d\tilde{\varepsilon} \nonumber \\
&&= \int_0^1 \left[
    \left( \varepsilon_{ij} - \varepsilon_{ik}\varepsilon_{kj}\tilde{\varepsilon} \right) 
    \sigma^{\rm (ext)}_{ij} \right] V d\tilde{\varepsilon}
  + \mathcal{O}(\boldsymbol{\varepsilon}^3),
\label{Eq1-11}
\end{eqnarray}
where $(1+\tilde{\varepsilon}\boldsymbol{\varepsilon})^{-1}$ has been expanded to second-order
terms. $V$ is the volume during the deformation process, i.e.,
\begin{eqnarray}
  V\left( \tilde{\varepsilon}\boldsymbol{\varepsilon} \right) 
  &=& V_0 \det \left[ \tilde{\varepsilon}\boldsymbol{\varepsilon} \right]   \nonumber \\
  &=& V_0 \left[ 1+ \varepsilon_{ii} \tilde{\varepsilon}
    + \frac{1}{2}\left( \varepsilon_{ii}\varepsilon_{jj}-\varepsilon_{ij}\varepsilon_{ji} \right)
    \tilde{\varepsilon}^2 \right] + \mathcal{O}(\boldsymbol{\varepsilon}^3), \nonumber \\
\label{Eq1-12}
\end{eqnarray}
where $V_0$ is the unstrained volume. Equation~(\ref{Eq1-11}) then becomes
\begin{eqnarray}
  &&W^{\rm (ext)} \nonumber \\
  &&= \int_0^1 \left[ 
    \left( \varepsilon_{ij} - \varepsilon_{ik}\varepsilon_{kj}\tilde{\varepsilon} \right) 
    \sigma^{\rm (ext)}_{ij} \right] 
  V_0 \left( 1+ \varepsilon_{ll}  \tilde{\varepsilon} \right) d\tilde{\varepsilon}
  + \mathcal{O}(\boldsymbol{\varepsilon}^3) \nonumber \\
  &&= V_0\left( \varepsilon_{ij} - \frac{1}{2}\varepsilon_{ik}\varepsilon_{kj}
  + \frac{1}{2}\varepsilon_{ij}\varepsilon_{kk} \right) \sigma^{\rm (ext)}_{ij} 
  + \mathcal{O}(\boldsymbol{\varepsilon}^3).
\label{Eq1-13}
\end{eqnarray}
Equation~(\ref{Eq1-13}) is applicable for any pre-applied constant stress. For the
special case of hydrostatic pressure $P_0$, $\boldsymbol{\sigma}^{\rm (ext)}_{ij}=-P_0\delta_{ij}$ 
and $W^{\rm (ext)}$ reduces to the conventional form of volume work (with $\Delta V = V-V_0$):
\begin{eqnarray}
  W^{\rm (ext)} &=& -V_0 \left( \varepsilon_{ii} - \frac{1}{2}\varepsilon_{ik}\varepsilon_{ki}
  + \frac{1}{2}\varepsilon_{ii}\varepsilon_{kk} \right) P_0  \nonumber \\
  &=& -P_0 \Delta V.
\label{Eq1-14}
\end{eqnarray}

Substituting Eq.~(\ref{Eq1-13}) into Eq.~(\ref{Eq1-5}), we obtain the actual work needed
to strain the system, i.e.,
\begin{eqnarray}
\Delta W &=& \Delta F-W^{\rm (ext)} \nonumber \\
&=& \Delta F- V_0\left( \varepsilon_{ij} - \frac{1}{2}\varepsilon_{ik}\varepsilon_{kj}
+\frac{1}{2}\varepsilon_{ij}\varepsilon_{kk} \right) \sigma^{\rm (ext)}_{ij} \nonumber \\
\label{Eq1-15}
\end{eqnarray}
up to second order of ${\bm \varepsilon}$, which determines the stability and elastic
coefficients of the system. When the initial state is equilibrated by
$\boldsymbol{\sigma}^{\rm (ext)}=\boldsymbol{\sigma}^{\rm (0)}$ (here a different symbol
$\boldsymbol{\sigma}^{\rm (0)}$ is used to emphasize that $\boldsymbol{\sigma}^{\rm (0)}$ 
stabilizes the initial state, while $\boldsymbol{\sigma}^{\rm (ext)}$ could be any
external stress under which the initial state is not necessarily stable),
\begin{equation}
\frac{\partial \Delta W}{\partial \varepsilon_{ij}}
=\left. \frac{\partial F}{\partial \varepsilon_{ij}} \right |_{\boldsymbol{\varepsilon}=0}
  - V_0\sigma^{\rm (0)}_{ij}=0,
\label{Eq1-16}
\end{equation}
which gives
\begin{equation}
  \sigma^{\rm (0)}_{ij} = \left. \frac{1}{V_0} \frac{\partial F}{\partial \varepsilon_{ij}}
  \right |_{\boldsymbol{\varepsilon}=0}.
\label{Eq1-17}
\end{equation}
Under this external stress $\boldsymbol{\sigma}^{\rm (0)}$, a Gibbs-type free energy
$G$ can be defined by requiring
\begin{equation}
\Delta G = \Delta W (\boldsymbol{\sigma}^{\rm (0)}).
\label{Eq1-17b}
\end{equation}
A solution to Eq.~(\ref{Eq1-17b}) is
\begin{equation}
  G= F+P_0V_0 -\left( \varepsilon_{ij} - \frac{1}{2}\varepsilon_{ik}\varepsilon_{kj}
  + \frac{1}{2}\varepsilon_{ij}\varepsilon_{kk} \right) \sigma^{\rm (0)}_{ij} V_0,  
\label{Eq1-17c}
\end{equation}
where a constant $P_0V_0$ is added to make it consistent with the standard definition
of Gibbs free energy when $\boldsymbol{\sigma}^{\rm (0)}$ is isotropic [i.e., when
$\sigma^{\rm (0)}_{ij} = -P_0 \delta_{ij}$; see Eq.~(\ref{Eq2-1})]. $G$ can be expanded as
\begin{eqnarray}
  G&=&F_0 +P_0V_0 +\frac{1}{2} \left. \frac{\partial^2 F}
  {\partial \varepsilon_{ij} \partial \varepsilon_{kl}} \right |_{\boldsymbol{\varepsilon}=0}
  \varepsilon_{ij}\varepsilon_{kl} \nonumber\\
  && +\frac{1}{2} \left. \frac{\partial F}{\partial \varepsilon_{ij}} \right |_{\boldsymbol{\varepsilon}=0}
  \left( \varepsilon_{ik}\varepsilon_{kj} - \varepsilon_{ij}\varepsilon_{kk}  \right)
  + \mathcal{O}(\boldsymbol{\varepsilon}^3)
\label{Eq1-18}
\end{eqnarray}
up to the second order of ${\bm \varepsilon}$, where there are no first-order terms
as in the conventional case of elastic response for a stable undeformed state. 
The elastic coefficients are thus defined as the second-order derivatives of $G$
(i.e., of the actual work done to deform the system) with respect to the strain
tensor components:
\begin{equation}
  C_{ijkl}= \frac{1}{V_0} \left. \frac{\partial^2 \Delta W}
  {\partial \varepsilon_{ij} \partial \varepsilon_{kl}} \right |_{\boldsymbol{\varepsilon}=0}
  = \frac{1}{V_0} \left. \frac{\partial^2 G}
  {\partial \varepsilon_{ij} \partial \varepsilon_{kl}} \right |_{\boldsymbol{\varepsilon}=0}.
\label{Eq1-19}
\end{equation}

To facilitate the calculation, we introduce an effective strain tensor [i.e.,
Eq.~(\ref{Eq1-20a})]
\begin{equation}
  \xi_{ij} = \varepsilon_{ij}- \frac{1}{2}\varepsilon_{ik}\varepsilon_{kj}
  + \frac{1}{2}\varepsilon_{ij}\varepsilon_{kk},
\label{Eq1-20}
\end{equation}
such that
\begin{equation}
  G = F + \left ( P_0 - \xi_{ij} \sigma^{\rm (0)}_{ij} \right ) V_0.
  \label{Eq_G}
\end{equation}
$F$ can be expanded to the second order of strain tensor as
\begin{eqnarray}
  F &=& F_0 + \left. \frac{\partial F}{\partial \xi_{ij}} \right |_{{\bm \xi}=0} \xi_{ij}
  +\frac{1}{2} \left. \frac{\partial^2 F}{\partial \xi_{ij} \partial \xi_{kl}} \right |_{{\bm \xi}=0}
  \xi_{ij}\xi_{kl} + \mathcal{O}({\bm \xi}^3) \nonumber \\
  &=& F_0  + \left. \frac{\partial F}{\partial \xi_{ij}} \right |_{{\bm \xi}=0}
  \left( \varepsilon_{ij} - \frac{1}{2}\varepsilon_{ik}\varepsilon_{kj} +
  \frac{1}{2}\varepsilon_{ij}\varepsilon_{kk} \right)
  \nonumber \\
  && +\frac{1}{2} \left. \frac{\partial^2 F}{\partial \xi_{ij} \partial \xi_{kl}} \right |_{{\bm \xi}=0}
  \varepsilon_{ij}\varepsilon_{kl} + \mathcal{O}(\boldsymbol{\varepsilon}^3).
  \label{Eq1-21}
\end{eqnarray}
Comparing it with
\begin{equation}
  F=F_0 + \left. \frac{\partial F}{\partial \varepsilon_{ij}} \right |_{\boldsymbol{\varepsilon}=0}
  \varepsilon_{ij} +\frac{1}{2} \left. \frac{\partial^2 F}
  {\partial \varepsilon_{ij} \partial \varepsilon_{kl}} \right |_{\boldsymbol{\varepsilon}=0}
  \varepsilon_{ij}\varepsilon_{kl} + \mathcal{O}(\boldsymbol{\varepsilon}^3),
\label{Eq1-22}
\end{equation}
we have
\begin{equation}
  \left. \frac{\partial F}{\partial \xi_{ij}} \right |_{{\bm \xi}=0}
  = \left. \frac{\partial F}{\partial \varepsilon_{ij}} \right |_{\boldsymbol{\varepsilon}=0},
\label{Eq1-23}
\end{equation}
and
\begin{eqnarray}
  && \left. \frac{\partial^2 F}{\partial \varepsilon_{ij} \partial \varepsilon_{kl}}
  \right |_{\boldsymbol{\varepsilon}=0} \varepsilon_{ij}\varepsilon_{kl} \label{Eq1-24} \\
  &&= \left. \frac{\partial F}{\partial \varepsilon_{ij}} \right |_{\boldsymbol{\varepsilon}=0}
  \left( - \varepsilon_{ik}\varepsilon_{kj} + \varepsilon_{ij}\varepsilon_{kk} \right)
  + \left. \frac{\partial^2 F}{\partial \xi_{ij} \partial \xi_{kl}} \right |_{{\bm \xi}=0}
  \varepsilon_{ij}\varepsilon_{kl}. \nonumber
\end{eqnarray}
Inserting Eq.~(\ref{Eq1-24}) into Eq.~(\ref{Eq1-18}) yields
\begin{equation}
  G=F_0 +P_0V_0 +\frac{1}{2} \left. \frac{\partial^2 F}{\partial \xi_{ij} \partial \xi_{kl}}
  \right |_{{\bm \xi}=0} \varepsilon_{ij}\varepsilon_{kl}.
\label{Eq1-25}
\end{equation}
From Eq.~(\ref{Eq1-19}) we then obtain an alternative formulation to determine the
elastic constants:
\begin{equation}
  C_{ijkl}= \frac{1}{V_0} \left. \frac{\partial^2 F}{\partial \xi_{ij} \partial \xi_{kl}}
  \right |_{{\bm \xi}=0}.
\label{Eq1-26}
\end{equation}
It is noted that $C_{ijkl}$ defined here corresponds to the stress-strain elastic
coefficient $B_{ijkl}$ defined by Birch \cite{pr809} and Wallace \cite{pr776} 
when $\boldsymbol{\sigma}^{\rm (0)}$ is isotropic (i.e., for the case of initial
isotropic pressure $P_0$), but in general cases they are not equivalent given that
$B_{ijkl}$ is lack of complete Voigt symmetry (generally $B_{ijkl} \neq B_{klij}$,
unless $\sigma^{\rm (0)}_{ij} = -P_0 \delta_{ij}$). \cite{pr776} 
Equations~(\ref{Eq1-19}) and (\ref{Eq1-26}) are two equivalent formulae to calculate the
elastic constants of stressed and unstressed systems. They are renumbered to
Eqs.~(\ref{Eq1-1c}) and (\ref{Eq1-26a}) above.

\section{First-order variation of $A_{\mathbf{K}}$ when $\bar{\psi}$ is unchanged under
  deformation}
\label{sec:appen_AK}

For the PFC free energy functional given in Eq.~(\ref{PFC-4}), after substituting
Eq.~(\ref{PFC-13}) for the expansion of $\psi$ and integrating over the system
volume $V$, the resulting free energy can be written in a general form as 
\begin{equation}
  \frac{1}{V} F(A_{\mathbf{K}},\mathbf{K}; \bar{\psi}, V)
  = f(\mathbf{K})A_{\mathbf{K}}^2+g(A_{\mathbf{K}}, \bar{\psi}),
\label{Eq3-1}
\end{equation}
for any crystalline phase. Here $f$ is a function of $\mathbf{K}$ and $g$ a function
of $A_{\mathbf{K}}$ and $\bar{\psi}$, with the detailed form of functions depending on
the specific phase. From Eq.~(\ref{PFC-15}), the equilibrium $\mathbf{K}$ is determined
by 
\begin{equation}
\left. \frac{d f(\mathbf{K})}{d \mathbf{K}} \right |_{\rm eq} = 0, 
\label{Eq3-2}
\end{equation}
which is independent of $\bar{\psi}$; i.e., $\mathbf{K}$ and the equilibrium lattice
constant are independent of the average atomic density (which is a drawback of this
PFC model that could be improved by e.g., incorporating nonlinear gradient terms
originated from high-order direct correlations in the free energy functional \cite{WangZL}).
On the other hand, $A_{\mathbf{K}}$ is determined from 
\begin{equation}
\frac{\partial F(A_{\mathbf{K}},\mathbf{K}; \bar{\psi}, V)/V}{\partial A_{\mathbf{K}}}=0,
\label{Eq3-3}
\end{equation}
and is thus generally a function of $\bar{\psi}$:
\begin{equation}
A_{\mathbf{K}}=h\left( f(\mathbf{K}), \bar{\psi}\right).
\label{Eq3-4}
\end{equation}
Note that $\mathbf{K}$ affects $A_{\mathbf{K}}$ via $f(\mathbf{K})$.

Now we consider the first-order variation in elastic response, i.e., $d A_{\mathbf{K}}$,
$d\mathbf{K}$, and $d\bar{\psi}$. For the scheme of Eq.~(\ref{Res5}), $\bar{\psi}$ 
remains invariant under a strain, i.e., $d\bar{\psi} =0$. $\mathbf{K}$ is changed
according to Eq.~(\ref{Strain-5}) as usual. Thus from Eq.~(\ref{Eq3-4}), the first-order
variation of $A_{\mathbf{K}}$ is given by
\begin{equation}
  d A_{\mathbf{K}}=
  \frac{\partial h\left( f(\mathbf{K}), \bar{\psi}\right)}{\partial f(\mathbf{K})} 
  \left. \frac{d f(\mathbf{K})}{d \mathbf{K}} \right |_{\rm eq} d{\bf \mathbf{K}}=0,
\label{Eq3-5}
\end{equation}
due to Eq.~(\ref{Eq3-2}). A similar result was also noticed by Wu {\it et al.} in 
examining some specific deformations in PFC. \cite{pre61601}
As a result, $A_{\mathbf{K}}$ is invariant at the first order when Eq.~(\ref{Res5})
is assumed.

\section{Procedure of numerical calculations}
\label{sec:appen_NDetails}

For a specific crystalline phase such as bcc, there is only one free parameter for
determining $\{\mathbf{K}\}$, i.e., the first-mode wave vector magnitude denoted as $q_0$. 
Substituting the $\psi$ expansion Eq.~(\ref{PFC-13}) into the PFC free energy functional
Eq.~(\ref{PFC-4}), $\mathcal{F}/V$ becomes a polynomial function of $q_0$, $\{A_{\mathbf{K}}\}$,
and $\bar{\psi}$ (see Eq.~(3) in Ref.~\onlinecite{prl35501}); so are its first- and
second-order derivatives. Their analytic forms can be obtained straightforwardly,
and utilized in the numerical minimization process described below.

First, to determine the equilibrium undeformed state, $\mathcal{F}$ is minimized
numerically with respect to variables $q_0$ and $\{A_{\mathbf{K}}\}$ (which are
degenerate) under the condition of fixed $\bar{\psi}=\bar{\psi}_{\rm unstrained}$ and
$V=V_{\rm unstrained}=V_0$, 
yielding the equilibrium values of $\mathbf{K}^{\rm (unstrained)}$
and $\{A_{\mathbf{K}}^{(0)}\}$. After then various strains are applied as follows:
Each one of and each pair of independent strain elements among
$\{\varepsilon_{11}, \varepsilon_{22}, \varepsilon_{33}, \varepsilon_{12},
\varepsilon_{13}, \varepsilon_{23}\}$ are chosen separately and assigned a nonzero
value that varies in a range from $-3\%$ to $3\%$, with the rest being kept zero.
Given each of the resulting strain tensor $\boldsymbol{\varepsilon}$, the
corresponding strained values of $\mathbf{K}^{\rm (strained)}$, $V_{\rm strained}$,
and $\bar{\psi}_{\rm strained}$ are calculated by Eqs.~(\ref{Strain-5}), (\ref{Strain-6}),
and (\ref{PFC-11}), respectively. Next, according to the scheme of Eq.~(\ref{Strain-7}),
given the values of $\mathbf{K}^{\rm (strained)}$ and $\bar{\psi}_{\rm strained}$
determined above, $\mathcal{F}$ is numerically minimized with respect to amplitudes
$\{A_{\mathbf{K}}\}$ (which are now assumed to be non-degenerate) through e.g., the
Newton-Raphson method, to give the value of strained-state free energy $F$ under
each assigned strain tensor $\boldsymbol{\varepsilon}$. 
The obtained data points of $F$ vs $\varepsilon_{ij}$
are then fitted into Eq.~(\ref{Eq1-22}) to give the first- and second-order derivatives
of $F(\boldsymbol{\varepsilon})$ with respect to $\varepsilon_{ij}$, which are
used to convert to the pressure $P_0$ [via Eq.~(\ref{Eq1-1b})] and elastic constants
$C_{ijkl}^{(\varepsilon)}$ [via Eq.~(\ref{Eq4-6})]. Similarly, $C_{ijkl}^{(E)}$ and $C_{ijkl}$
are calculated from the fitting of those strained-state $F$ data points to the
second-order expansions of $F$ vs $\mathbf{E}$ and $F$ vs $\boldsymbol{\xi}$ and
then the use of the corresponding elastic constant definitions Eq.~(\ref{Eq4-12})
and Eq.~(\ref{Eq1-26a}), respectively. Here the values of finite strain tensors
$\mathbf{E}$ and $\boldsymbol{\xi}$ are calculated from $\boldsymbol{\varepsilon}$
according to Eqs.~(\ref{Strain-4}) and (\ref{Eq1-20a}).


\begin{thebibliography}{99}

\bibitem{prl245701} K. R. Elder, M. Katakowski, M. Haataja, and M. Grant, Phys. Rev. Lett.
  {\bf 88}, 245701 (2002).

\bibitem{prl225504} P. Stefanovic, M. Haataja, and N. Provatas, Phys. Rev. Lett.
  {\bf 96}, 225504 (2006).

\bibitem{prb64107} K. R. Elder, N. Provatas, J. Berry, P. Stefanovic, and M. Grant,
  Phys. Rev. B {\bf 75}, 064107 (2007). 

\bibitem{pre21605} Z.-F. Huang, K. R. Elder, and N. Provatas, Phys. Rev. E {\bf 82}, 021605 (2010).

\bibitem{ap665} H. Emmerich, H. Lowen, R. Wittkowski, T. Gruhn, G. I. Toth, G. Tegze,
  and L. Granasy, Adv. Phys. {\bf 61}, 665 (2012).

\bibitem{prb6119} J. B. Collins and H. Levine, Phys. Rev. B {\bf 31}, 6119 (1985).

\bibitem{arms113} See, e.g., L. Q. Chen, Annu. Rev. Mater. Res. {\bf 32}, 113 (2002),
  and references therein.

\bibitem{prb35429} Z. R. Liu, H. J. Gao, L. Q. Chen, and K. J. Cho, Phys. Rev. B
  {\bf 68}, 035429 (2003).

\bibitem{pre51605} K. R. Elder and M. Grant, Phys. Rev. E {\bf 70}, 051605 (2004).

\bibitem{prb184110} J. Mellenthin, A. Karma, and M. Plapp, Phys. Rev. B {\bf 78}, 184110 (2008).

\bibitem{prl255501} D. Taha, S. K. Mkhonta, K. R. Elder, and Z.-F. Huang,
  Phys. Rev. Lett. {\bf 118}, 255501 (2017).

\bibitem{prb184107} K. A. Wu and A. Karma, Phys. Rev. B {\bf 76}, 184107 (2007).

\bibitem{pre31602} A. Jaatinen, C. V. Achim, K. R. Elder, and T. Ala-Nissila,
  Phys. Rev. E {\bf 80}, 031602 (2009).

\bibitem{prl35702} G. Tegze, L. Granasy, G. I. Toth, F. Podmaniczky, A. Jaatinen,
  T. Ala-Nissila, and T. Pusztai, Phys. Rev. Lett. {\bf 103}, 035702 (2009).

\bibitem{pre012405} S. Tang, Y.-M. Yu, J. C. Wang, J. J. Li, Z. J. Wang, Y. L. Guo,
  and Y. H. Zhou, Phys. Rev. E {\bf 89}, 012405 (2014).

\bibitem{prl15502} P. Y. Chan, G. Tsekenis, J. Dantzig, K. A. Dahmen, and N. Goldenfeld,
  Phys. Rev. Lett. {\bf 105}, 015502 (2010).

\bibitem{prb184109} M. Seymour, F. Sanches, K. R. Elder, and N. Provatas, Phys. Rev. B
  {\bf 92}, 184109 (2015).

\bibitem{pre22105} E. Alster, K. R. Elder, J. J. Hoyt, and P. W. Voorhees, Phys. Rev. E
  {\bf 95}, 022105 (2017).

\bibitem{prl045702} M. Greenwood, N. Provatas, and J. Rottler, Phys. Rev. Lett.
  {\bf 105}, 045702 (2010).

\bibitem{jpc205402} A. Jaatinen and T. Ala-Nissila, J. Phys.: Condens. Matter
  {\bf 22}, 205402 (2010).

\bibitem{jpc364102} K. A. Wu, M. Plapp, and P. W. Voorhees, J. Phys.: Condens. Matter
  {\bf 22}, 364102 (2010).

\bibitem{prm060801} E. Alster, D. Montiel, K. Thornton, and P. W. Voorhees, 
  Phys. Rev. Materials {\bf 1}, 060801 (2017).

\bibitem{pre53305} V. W. L. Chan, N. Pisutha-Arnond, and K. Thornton, Phys. Rev. E
  {\bf 91}, 053305 (2015).

\bibitem{npj15013} M. Lavrskyi, H. Zapolsky, and A. G. Khachaturyan, NPJ Comput. Mater.
  {\bf 2}, 15013 (2016).

\bibitem{prl35501} S. K. Mkhonta, K. R. Elder, and Z.-F. Huang, Phys. Rev. Lett.
  {\bf 111}, 035501 (2013).

\bibitem{prl205502} S. K. Mkhonta, K. R. Elder, and Z.-F. Huang, Phys. Rev. Lett.
  {\bf 116}, 205502 (2016).

\bibitem{prb125408} K. A. Wu and P. W. Voorhees, Phys. Rev. B {\bf 80}, 125408 (2009).

\bibitem{pre61601} K. A. Wu, A Adland, and A. Karma, Phys. Rev. E {\bf 81}, 061601 (2010).

\bibitem{pre11602} K. R. Elder, Z.-F. Huang, and N. Provatas, Phys. Rev. E
  {\bf 81}, 011602 (2010).

\bibitem{prb214105} C. Huter, M. Friak, M. Weikamp, J. Neugebauer, N. Goldenfeld,
  B. Svendsen, and R. Spatschek, Phys. Rev. B {\bf 93}, 214105 (2016).

\bibitem{prb14103} N. Pisutha-Arnond, V. W. L. Chan, K. R. Elder, and K. Thornton,
  Phys. Rev. B {\bf 87}, 014103 (2013).

\bibitem{pr776} D. C. Wallace, Phys. Rev. {\bf 162}, 776 (1967).

\bibitem{prb423} J. R. Ray, Phys. Rev. B {\bf 40}, 423 (1989).

\bibitem{jasa348} R. N. Thurston, J. Acoust. Soc. Am. {\bf 37}, 348 (1965).

\bibitem{pm3977} M. I. Mendelev, S. Han, D. J. Srolovitz, G. J. Ackland, D. Y. Sun,
  and M. Asta, Philos. Mag. {\bf 83}, 3977 (2003).

\bibitem{pr809} F. Birch, Phys. Rev. {\bf 71}, 809 (1947).

\bibitem{WangZL} Z.-L. Wang, Z. R. Liu, and Z.-F. Huang, preprint.

\bibitem{prb165421} Z.-F. Huang and K. R. Elder, Phys. Rev. B {\bf 81}, 165421 (2010).

\bibitem{pre32411} V. Heinonen, C. V. Achim, K. R. Elder, S. Buyukdagli, and T. Ala-Nissila,
  Phys. Rev. E {\bf 89}, 032411 (2014).

\bibitem{prb12627} J. Wang, J. Li, S. Yip, S. Phillpot, and D. Wolf, Phys. Rev. B {\bf 52},
  12627 (1995).

\bibitem{pma2827} J. W. Morris Jr. and C. R. Krenn, Philos. Mag. A {\bf 80}, 2827 (2000).

\end{thebibliography}
\end{document}